\documentclass[11pt]{article}
\usepackage{fullpage}
\usepackage{graphicx}
\usepackage{latexsym,amsmath,amssymb,amsthm,epic,eepic,multirow}
\usepackage{natbib}
\usepackage{multicol}
\usepackage{multirow}
\usepackage{subcaption} 
\usepackage{amsfonts}
\usepackage{natbib}
\usepackage{algorithm}
\usepackage[noend]{algpseudocode}
\usepackage{mdframed}
\usepackage{tikz}
\usetikzlibrary{positioning, calc}
\usepackage{hyperref}
\usepackage{listings} 
\usepackage{color} 
\usepackage{bbm}
\usepackage{relsize}
\usepackage{array}
\usepackage{mathtools}
\newcolumntype{P}[1]{>{\centering\arraybackslash}p{#1}}
\newcolumntype{R}[1]{>{\raggedleft\arraybackslash}p{#1}}

\definecolor{upmaroon}{rgb}{0.48, 0.07, 0.07}
\definecolor{royalazure}{rgb}{0.0, 0.22, 0.66}
\definecolor{pakistangreen}{rgb}{0.0, 0.4, 0.0}

\newcommand{\PP}{\mathbb{P}}
\newcommand{\EE}{\mathbb{E}}

\theoremstyle{definition}
\newtheorem{theorem}{Theorem}
\newtheorem{proposition}{Proposition}
\newtheorem{lemma}{Lemma}

\newtheorem{definition}{Definition}

\newtheoremstyle{dotless}{}{}{}{}{\bfseries}{}{ }{}
\theoremstyle{dotless}
\newtheorem{assa}{}

\usepackage{algorithm}
\usepackage{algpseudocode}

\algnewcommand{\Inputs}[1]{%
  \State \textbf{Inputs:}
  \Statex \hspace*{\algorithmicindent}\parbox[t]{.8\linewidth}{\raggedright #1}
}
\algnewcommand{\Initialize}[1]{%
  \State \textbf{Initialize:}
  \Statex \hspace*{\algorithmicindent}\parbox[t]{.8\linewidth}{\raggedright #1}
}
\algnewcommand{\Returns}[1]{%
  \State \textbf{Returns:}
  \Statex \hspace*{\algorithmicindent}\parbox[t]{.8\linewidth}{\raggedright #1}
}

\usepackage{color}

\let\originalleft\left
\let\originalright\right
\renewcommand{\left}{\mathopen{}\mathclose\bgroup\originalleft}

\renewcommand{\right}{\aftergroup\egroup\originalright}
\makeatletter
\newcommand{\leqnomode}{\tagsleft@true}
\newcommand{\reqnomode}{\tagsleft@false}
 
\makeatother
\begin{document}
\title{Ancestor regression in structural vector autoregressive models} 

\author{Christoph Schultheiss, Markus Ulmer, and Peter B\"uhlmann\\
Seminar for Statistics, ETH Z\"urich}

\maketitle

\begin{abstract}
We present a new method for causal discovery in linear structural vector autoregressive models. We adapt an idea designed for independent observations to the case of time series while retaining its favorable properties, i.e., explicit error control for false causal discovery, at least asymptotically. We apply our method to several real-world bivariate time series datasets and discuss its findings which mostly agree with common understanding. 

The arrow of time in a model can be interpreted as background knowledge on possible causal mechanisms. Hence, our ideas could be extended to incorporating different background knowledge, even for independent observations.
\end{abstract}


\section{Introduction}
Real-world datasets often exhibit a time structure violating the i.i.d.\ assumption widely used in causal discovery and beyond. Such data can be modelled with (structural) vector autoregressive models, i.e., using past and current observations of the time series as predictors. While the time dependence implies certain difficulties in estimation, it also offers some advantages because a predictor cannot causally affect other variables that represent earlier time points. With independent innovation terms, identifiability guarantees as for fully independent observations can be found under similar structural assumptions, see \cite{peters2013causal}. 

\subsection{Our contribution}
In this work, we extend the recent development on ancestor regression by \cite{schultheiss2023ancestor} to the case of multivariate time series with linear causal relations, both instantaneous and lagged. The time dependence between the observations poses technical challenges to ensure the asymptotic guarantees. Further, to obtain error control among the lagged effects, we show how to choose the right adjustment sets for different time lags. Given the amount of time series data encountered in applications, we feel that this extension is of significant practical use; see also the empirical demonstration in Section \ref{real-data}.

\subsection{Structural vector autoregressive model}
Let us denote the observed time series by $x_{t,j}$ for $t=1, \ldots, T$ and $j=1,\ldots,d$. At time $t$ the variables are collected to the vector $\mathbf{x}_t = (x_{t,1},\ldots ,x_{t,d})^T$. We assume strictly stationary time series, i.e., the probabilistic behavior is the same for every $t$. We say the time series follows a structural vector autoregressive (SVAR) model of order $p$ if
\begin{equation}\label{eq:svar}
\mathbf{x}_t = \sum_{\tau = 0}^p \mathbf{B}_{\tau} \mathbf{x}_{t-\tau} + \boldsymbol{\epsilon}_{t}, \quad \text{where} \quad \boldsymbol{\epsilon}_t = (\epsilon_{t,1},\ldots ,\epsilon_{t,d})^T.
\end{equation}
We make the following assumptions:
\begin{assa}\label{ass:iid}
The $\epsilon_{t,j}$ are centered, independent over both, $t$ and $j$, and identically distributed over $t$.
\end{assa}
\begin{assa}\label{ass:dag}
The instantaneous effects in $\mathbf{B}_0$ imply an acyclic structure.
\end{assa}

Instantaneous effects are often assumed to represent processes that happen at a higher frequency than the observation sampling frequency. As such it is not clear that they must be acyclic. However, the assumption is standard when extending causal discovery algorithms to applications on time series data, see, e.g., \cite{peters2013causal}. We follow this proposal and take the assumption as a given. However, as cyclic effects from high-frequency processes can be attributed to unobserved variables, the impact is implicitly covered by our simulation study in Section \ref{network-sim}.

\ref{ass:dag} implies that $\mathbf{B}_0$ corresponds to a row- and column-permuted lower triangular matrix. Therefore, the eigenvalues of $I-\mathbf{B}_0$ are all $1$ and it is invertible. Hence, we get the equivalent form of our model
\begin{equation}\label{eq:var}
\mathbf{x}_t = \left(I-\mathbf{B}_0\right)^{-1}\left(\sum_{\tau = 1}^p \mathbf{B}_{\tau} \mathbf{x}_{t-\tau} + \boldsymbol{\epsilon}_{t}\right) \coloneqq \sum_{\tau = 1}^p \tilde{\mathbf{B}}_{\tau} \mathbf{x}_{t-\tau} + \boldsymbol{\xi}_{t},
\end{equation}
where $\boldsymbol{\xi}_{t}$ has correlated components but is independent over time. Let $\mathbf{x}_{t\_p+1}$ with $p \geq 0$ be the time series in time range $t$ to $t-p$ flattened out to a vector in $\mathbb{R}^{d\left(p + 1\right)}$. Also $\boldsymbol{\xi}_{t\_p+1}$ are flattened versions of $\boldsymbol{\xi}_{t}$ patched with zeros, i.e.,
\begin{equation*}
\mathbf{x}_{t\_p+1} = \begin{pmatrix}
\mathbf{x}_{t} \\
\mathbf{x}_{t-1}\\
\vdots \\
\mathbf{x}_{t-p}
\end{pmatrix} \in \mathbb{R}^{d\left(p+1\right)} \quad \text{and} \quad \boldsymbol{\xi}_{t\_p+1} = \begin{pmatrix}
\boldsymbol{\xi}_{t} \\
0\\
\vdots \\
0
\end{pmatrix} \in \mathbb{R}^{d\left(p+1\right)}.
\end{equation*}
With these flattened versions, we can rewrite \eqref{eq:var} as an order $1$ model
\begin{equation}\label{eq:comp}
\mathbf{x}_{t\_p} = \begin{pmatrix}
\tilde{\mathbf{B}}_{1} & \tilde{\mathbf{B}}_{2} & \ldots & \tilde{\mathbf{B}}_{p- 1}&  \tilde{\mathbf{B}}_{p}\\
I & 0 & \ldots & 0 & 0 \\
0 & I & \ldots & 0 & 0 \\
\vdots & & \ddots & \vdots & \vdots \\
0 & 0 & \ldots & I & 0
\end{pmatrix} \mathbf{x}_{t-1\_p} + \boldsymbol{\xi}_{t\_p} \coloneqq \tilde{\mathbf{B}} \mathbf{x}_{t-1\_p} + \boldsymbol{\xi}_{t\_p};
\end{equation}
see, e.g., \cite[Chapter~2]{lutkepohl2005new}. We additionally require
\begin{assa}\label{ass:stable}
The process $\mathbf{x}_t$ is stable, i.e., for $\tilde{\mathbf{B}}$ as in \eqref{eq:comp}, $\text{det}\left(I - \tilde{\mathbf{B}}s\right)\neq 0$ if $\left\vert s \right\vert \leq 1$.
\end{assa}
This implies strict stationarity if the process is initialized correctly or has run for an infinite time.

The setting mostly corresponds to the one in \cite{hyvarinen2010estimation} which extends the LiNGAM method from \cite{shimizu2006linear} for linear structural equation models in the i.i.d.\ setting to the time series case.

Let the $\tau$-lagged causal ancestors of $j$, $\text{AN}^{\tau}\left(j\right)$ be all $k$ for which there exists a directed path from $x_{t-\tau,k}$ to $x_{t,j}$ for all $t$ in the full causal graph. Analogously, we say $k \in \text{PA}^{\tau}\left(j\right) \iff j \in \text{CH}^{\tau}\left(k\right)$ to denote parents and children if there is an edge from $x_{t-\tau,k}$ to $x_{t,j}$ for all $t$. For $\tau = 0$, $\text{AN}^{\tau = 0}(j)$ are the instantaneous ancestors of $j$. As the $\mathbf{B}_{\tau}$ are time-invariant, we also omit the time index for ancestors, children, and parents.

\subsection{Identifying $AN^\tau$ via ancestor regression}
For the case of linear causal relations in i.i.d.\ data, the recent development in \cite{schultheiss2023ancestor} provides asymptotic type I error guarantees for detecting any covariate's ancestors. The method revolves around the following key observation: Assume that a set of variables $x_k$, $k \in \left\{1,\ldots,p\right\}$ is connected by linear causal relations (plus additive noise), and we are interested in the causal ancestors of a given $x_j$. Then, we can use least squares regression with response variable $f\left(x_j\right)$, where $f\left(\cdot\right)$ is a nonlinear function, such as $f\left(x_j\right) = x_j^3$, and all $x_k$ as predictors, including $x_j$ itself: 
\begin{equation*}
    f(x_j)\ \text{versus}\ x_j, \{x_k; k \neq j\}\ \text{with least squares}.
\end{equation*}
The resulting least squares coefficients are (in population) $0$ for all non-ancestors, while they are - up to few counter-examples - non-zero for the ancestors. For a discussion of such counter-examples, see Section 2.2 in \cite{schultheiss2023ancestor}, we adapt this to time series in Theorem \ref{SVAR:theo:adv}. The most important exception is Gaussian noise. If there are adjacent variables with Gaussian noise, several causal orders could imply the same multivariate distribution. Due to the type I guarantees, ancestor regression cannot detect an ancestral relation in such cases and remains conservative. Accordingly, the related LiNGAM algorithm \citep{shimizu2006linear} allows for at most one Gaussian noise term by assumption - otherwise it becomes inconsistent.

Based on the observation above, one can identify the ancestors, or at least some of them, using a simple least squares regression. We will show here how this method can be extended to the related SVAR \eqref{eq:svar}.

Let
\begin{equation}\label{eq:Atau}
\mathbf{x}_t \coloneqq \mathbf{A}_{\tau} \mathbf{x}_{t-\left(\tau + 1\right)\_p} + \boldsymbol{\xi}^{\tau}_{t} \quad \text{with} \quad \mathbf{x}_{t-\left(\tau + 1\right)\_p} \perp \boldsymbol{\xi}^{\tau}_{t}.
\end{equation}
Here and forthcoming, we use $\perp$ to denote statistical independence. This means that we regress out the contribution of the observations from $\tau +1$ to $\tau + p$ time steps before. Due to the independence of the innovation terms, such an independent residual can always be found. Hence, $\boldsymbol{\xi}^{\tau}_{t}$ are also the corresponding least squares residuals. For $\tau = 0$ and
using \eqref{eq:comp}, we obtain:
\begin{equation*}
\mathbf{A}_0 = \begin{pmatrix}
\tilde{\mathbf{B}}_{1} & \tilde{\mathbf{B}}_{2} & \ldots & \tilde{\mathbf{B}}_{p- 1}&  \tilde{\mathbf{B}}_{p}
\end{pmatrix} \quad \text{and} \quad \boldsymbol{\xi}^{0}_{t} = \boldsymbol{\xi}_{t}.
\end{equation*}
Define further
\begin{align*}
z_{t,k} & \coloneqq \xi_{t,k} - \boldsymbol{\xi}_{t,-k}^{\top}\boldsymbol{\gamma}_k, \quad \text{where}\quad
\boldsymbol{\gamma}_k \coloneqq \underset{\mathbf{b} \in \mathbb{R}^{d-1}}{\text{argmin}}\EE\left[\left(\xi_{t,k} - \boldsymbol{\xi}_{t,-k}^{\top} \mathbf{b}\right)^2\right] = \EE\left[\boldsymbol{\xi}_{t,-k}\boldsymbol{\xi}_{t,-k}^{\top}\right]^{-1}\EE\left[\boldsymbol{\xi}_{t,-k}\xi_{t,k}\right],
\end{align*}
i.e., the least squares residual of regressing one $\xi_{t,k}$ against all others, or, equivalently, the residual of regressing one $x_{t,k}$ against all others and $\mathbf{x}_{t-1\_p}$.
\begin{theorem}\label{theo:bols}
Assume that $\mathbf{x}_t$ follows the model \eqref{eq:svar} with assumptions \ref{ass:iid} to \ref{ass:stable}. Consider the ordinary least squares regression $f\left(\xi^{\tau}_{t,j}\right)$ versus $\boldsymbol{\xi}_{t-\tau}$ and denote the according corresponding OLS parameter by 
\begin{equation*}
\boldsymbol{\beta}^{f,j,\tau} \coloneqq \EE\left[\boldsymbol{\xi}_{t-\tau}\boldsymbol{\xi}_{t-\tau}^\top\right]^{-1}\EE\left[\boldsymbol{\xi}_{t-\tau}f\left(\xi^{\tau}_{t,j}\right)\right]
\end{equation*}
and assume that it exists. Then,
\begin{equation*}
\beta^{f,j,\tau}_k =\EE\left[z_{t-\tau,k}f\left(\xi^{\tau}_{t,j}\right)\right]/\EE\left[z_{t-\tau,k}^2\right] = 0 \quad \forall k \not \in \text{AN}^{\tau}\left(j\right).
\end{equation*}
\end{theorem}
All, $\xi^{\tau}_{t,j}$ and $\boldsymbol{\xi}_{t-\tau}$, are residuals of a model using $\mathbf{x}_{t-\left(\tau + 1\right)\_p}$ as predictors and do not depend on time before $t-\tau$.

For $\tau = 0$ this construction corresponds to i.i.d.\ ancestor regression \citep{schultheiss2023ancestor} applied to $\boldsymbol{\xi}_{t}$ which follow an acyclic linear structure equation model as argued in \cite{hyvarinen2010estimation}.

Of particular interest is the reverse statement of Theorem \ref{theo:bols}, namely whether $\beta_k^{f,j,\tau}$ is non-zero for $k \in \text{AN}^{\tau}\left(j\right)$ for a nonlinear function $f(\cdot)$. While this is typically true, there are some adversarial cases as discussed next.
\subsubsection*{Adversarial setups}
There can be cases where $\beta_k^{f,j,\tau}=0$ although $k \in \text{AN}^\tau \left(j\right)$. For some data generating mechanisms, this happens regardless of the choice of function $f\left(\cdot\right)$. We want to characterize these cases. For instantaneous effects, these exceptions are - non-surprisingly - in close correspondence to the i.i.d.\  case. For lagged effects, for which the causal paths do not start with an instantaneous effect, it is related to faithfulness \citep[Chapter~2.3.3]{spirtes2000causation}. If the target is independent of its lagged ancestor due to the cancellation of terms, we cannot detect this relationship.

For the precise statements, we introduce some notation.
Denote the Markov boundary of $x_{t,k}$ by $ \text{MA}\left(k\right)$ and get the corresponding vector as
\begin{equation*}
\mathbf{x}_{t,\text{MA}\left(k\right)} \coloneqq \left\{x_{t,l}: l \in \text{CH}^{0}\left(k\right); \quad x_{t-\tau',l}: l \in \text{PA}^{\tau'}\left(k\right); \quad  x_{t-\tau',l} : l \in \text{PA}^{\tau'}\left(m\right) \ \text{and} \ m \in \text{CH}^{0}\left(k\right)\right\},
\end{equation*}
not including $x_{t,k}$ itself. This matches the classical definition of children, parents, and children's other parents but is restricted to the observed $\mathbf{x}_{t\_p+1}$. Now let 
\begin{equation*}
\text{CH}^{0,\tau \rightarrow j}\left(k\right) \coloneqq \text{CH}^{0}\left(k\right) \cap \text{AN}^{\tau}\left(j\right),
\end{equation*}
i.e., if $l\in\text{CH}^{0,\tau \rightarrow j}\left(k\right)$, there is a path $x_{t,k} \rightarrow x_{t,l} \rightarrow \ldots \rightarrow x_{t+\tau, j}$ for all $t$.
Then, we call $\text{MA}^{\tau \rightarrow j}\left(k\right)$ the restricted Markov boundary and get the corresponding vector as
\begin{align*}
\mathbf{x}_{t,\text{MA}^{\tau \rightarrow j}\left(k\right)} \coloneqq \Big\{& x_{t,l}: l \in \text{CH}^{0,\tau \rightarrow j}\left(k\right); \quad x_{t-\tau',l}: l \in \text{PA}^{\tau'}\left(k\right);\\
& x_{t-\tau',l} : l \in \text{PA}^{\tau'}\left(m\right) \ \text{and} \ m \in \text{CH}^{0,\tau \rightarrow j}\left(k\right)\Big\},
\end{align*}
again, not including $x_{t,k}$ itself, i.e., only children with a directed path to $x_{t+\tau, j}$ are considered.

\begin{theorem}\label{SVAR:theo:adv}
Let $k \in \text{AN}^\tau\left(j\right)$. 
Then, 
\begin{equation*}
\beta_k^{f,j,\tau}=0 \quad \forall f\left(\cdot\right) \quad \text{if and only if} \quad E\left[x_{t,k} \mid \xi^{\tau}_{t + \tau, j}\right] =  E\left[\mathbf{x}_{t,\text{MA}^{\tau \rightarrow j}\left(k\right)}^\top \boldsymbol{\gamma}^{\tau \rightarrow j,k} \mid \xi^{\tau}_{t + \tau, j}\right],
\end{equation*}
where $\boldsymbol{\gamma}^{\tau \rightarrow j,k}$ is the least squares parameter for regressing $x_{t,k}$ versus $\mathbf{x}_{t,\text{MA}^{\tau \rightarrow j}}$. This implies 
\begin{equation*}
x_{t,k} \perp x_{t + \tau, j} \mid \mathbf{x}_{t,\text{MA}^{\tau \rightarrow j}\left(k\right)}.
\end{equation*}
In particular,
\begin{equation*}
\beta_k^{f,j, \tau}=0 \quad \forall f\left(\cdot\right) \quad \text{if} \quad E\left[x_{t,k} \mid \mathbf{x}_{t,\text{MA}^{\tau \rightarrow j}\left(k\right)}\right] =  \mathbf{x}_{t,\text{MA}^{\tau \rightarrow j}\left(k\right)}^\top \boldsymbol{\gamma}^{\tau \rightarrow j,k} \quad \text{and} \quad x_{t,k} \perp x_{t + \tau, j} \mid \mathbf{x}_{t,\text{MA}^{\tau \rightarrow j}\left(k\right)}.
\end{equation*}
\end{theorem}
We provide some intuition for the necessary conditional independence.
It is always fulfilled for instantaneous effects as the restricted Markov boundary blocks all mediating or confounding paths. Hence, the sufficient condition boils down to the first equality only which is in correspondence to Theorem 3 of \cite{schultheiss2023ancestor} and the same corresponding examples apply, i.e., Gaussian error terms and identical contribution between predictor and noise, see there for details.

For a lagged effect, if there are paths from $x_{t,k}$ to $x_{t + \tau, j}$ that do not go through some other $x_{t,l}$, conditional independence only holds if those paths cancel each other out, i.e., the data is unfaithful. If there are only such paths, $\text{CH}^{0,\tau \rightarrow j}\left(k\right) = \emptyset$. This implies that $\text{MA}^{\tau \rightarrow j}\left(k\right)$ contains only ancestors such that the conditional expectation is indeed linear. Thus, in that case, unfaithfulness becomes sufficient and necessary.

Paths that go through some other $x_{t,l}$, i.e., starting with an immediate effect are a composition of the two cases above. The effects are detectable if both, the effect from $x_{t,k}$ to $x_{t,l}$ and from $x_{t,l}$ to $x_{t+\tau,k}$ are as well.

\section{Estimation from data and asymptotics}
Based on Theorem \ref{theo:bols}, we suggest testing for $\beta^{f,j,\tau}_k \neq 0$
in order to detect some or even all $\tau$-lagged causal ancestors of $x_{t,j}$. Doing so for all $k$ requires nothing more than fitting a multiple linear model and using its corresponding z-tests for individual covariates. Notably, if we are interested in $\text{AN}^\tau\left(j\right)$ for several values of $\tau$, we also consider several OLS regressions.

Let
\begin{equation*}
\mathbf{x}_{r:s\_p+1}=\begin{pmatrix}
\mathbf{x}_{r\_p+1} & \mathbf{x}_{r+1\_p+1} & \ldots & \mathbf{x}_{s\_p+1}
\end{pmatrix}^\top
\end{equation*}
for some $p+1\leq r\leq s \leq T$ be a matrix containing predictors at all lags for several time steps. Of course, this matrix has multiple entries corresponding to the same observation. Accordingly,
\begin{equation*}
\mathbf{x}_{r:s,j} = \begin{pmatrix}
x_{r,j} & x_{r+1,j} & \ldots & x_{s,j}.
\end{pmatrix}^\top
\end{equation*}
We get the least squares residuals' estimates for the residuals of interest.
\begin{align*}
\hat{\mathbf{z}}_{k}&=\hat{\mathbf{z}}_{p+1:T,k} \quad &&\text{the least squares residual of} \quad \mathbf{x}_{p+1:T,k} \quad \text{versus} \quad \mathbf{x}_{p+1:T,-k} \quad \text{and} \quad  \mathbf{x}_{p:T-1 \_p},\\
\hat{\boldsymbol{\xi}}^{\tau}_{k}&=\hat{\boldsymbol{\xi}}^{\tau}_{p+1+\tau:T,k} \quad &&\text{the least squares residual of} \quad \mathbf{x}_{p+1+\tau:T,k} \quad \text{versus} \quad \mathbf{x}_{p:T-\tau -1 \_p},\\
\hat{\boldsymbol{\xi}}^{\tau} &= \begin{pmatrix}
\hat{\boldsymbol{\xi}}^{\tau}_1 & \ldots & \hat{\boldsymbol{\xi}}^{\tau}_d
\end{pmatrix}^\top , \\
\hat{\boldsymbol{\xi}} & = \hat{\boldsymbol{\xi}}^{0}.
\end{align*}
Then, we calculate the following estimates
\begin{equation}\label{SVAR:eq:hat-def}
\begin{aligned}
\hat{\beta}^{f,j,\tau}_k & \coloneqq \hat{\mathbf{z}}_{p+1:T-\tau,k}^\top f\left(\hat{\boldsymbol{\xi}}^{\tau}_{j}\right)/\left\Vert\hat{\mathbf{z}}_{p+1:T-\tau,k}\right\Vert_2^2\\
\hat{\sigma}^2 & \coloneqq \dfrac{\left\Vert f\left(\hat{\boldsymbol{\xi}}^{\tau}_{j}\right) - \hat{\boldsymbol{\xi}}_{p+1:T-\tau,k}\hat{\boldsymbol{\beta}}^{f,j,\tau}\right\Vert_ 2^2}{T- d - \left(p + \tau\right)} \quad \text{and}\\
\widehat{\text{var}}\left(\hat{\beta}^{f,j,\tau}_k\right) & \coloneqq \hat{\sigma}^2/\left\Vert\hat{\mathbf{z}}_{p+1:T-\tau,k}\right\Vert_2^2,
\end{aligned}
\end{equation}
where $f\left(\cdot\right)$ is meant to be applied elementwise in $f\left(\hat{\boldsymbol{\xi}}^{\tau}_{j}\right)$. These are the classical least squares estimates for the given predictors and targets. There are $d$ covariates and $T-p-\tau$ observations can be used, hence the given normalization for the variance estimate. We obtain the test statistics
$s_k^{j,\tau} \coloneqq \dfrac{\hat{\beta}^{f,j,\tau}_k }{\surd{\widehat{\text{var}}\left(\hat{\beta}^{f,j,\tau}_k\right)}}$ for which we establish asymptotic normality under the null below. Therefore, we suggest testing the null hypothesis 
\begin{equation*}
H_{0,k^\tau\rightarrow j}: \ k \not \in \text{AN}^{\tau}\left(j\right)
\end{equation*}
with the p-value
\begin{equation}\label{SVAR:eq:z-test}
p_k^{j,\tau} = 2 \left\{1 - \Phi \left(\vert s_k^{j,\tau}\vert\right)\right\},
\end{equation}
where $\Phi \left(\cdot\right)$ denotes the cumulative distribution function of the standard normal distribution. 
We summarize the procedure in Algorithm \ref{alg:SVARAnc}. Note that we sometimes suppress explicit time indexing for easier readability.

To control the asymptotic behavior of these estimates, we make additional assumptions on $f\left(\cdot\right)$.
\begin{assa}\label{ass:fcond}
The function $f\left(\cdot\right)$ has the following properties
\begin{equation*}
\EE\left[ f\left(\xi^{\tau}_{t,j}\right)^2\right] < \infty \quad \text{and} \quad \exists \delta > 0 \quad \text{such that}\quad \EE\left[\left\vert \epsilon_{t,k}f\left(\xi^{\tau}_{t+\tau,j}\right)\right\vert^{1+\delta}\right] < \infty \quad \forall k.
\end{equation*}
Also, it is differentiable everywhere, and its derivative $f'\left(\cdot\right)$ has the following properties
\begin{equation*}
\exists \delta > 0 \quad \text{such that}\quad \EE\left[\left\vert f'\left(\xi^{\tau}_{t,j}\right)\right\vert^{2+\delta}\right] < \infty \quad \text{and} \quad \EE\left[\left\vert \epsilon_{t,k}f'\left(\xi^{\tau}_{t+\tau,j}\right)\right\vert^{1+\delta}\right] < \infty \quad \forall k.
\end{equation*}
\end{assa}
For monomials of the form
\begin{equation*}
f\left(x\right) = \text{sign}\left(x\right)\left\vert x\right\vert^\alpha, \quad \alpha > 1,
\end{equation*}
the moment conditions on $f\left(\cdot\right)$ imply those on $f'\left(\cdot\right)$. We use these functions by default.

\begin{theorem}\label{SVAR:theo:norm}
Let $\mathbf{x}_t$ follow an SVAR \eqref{eq:svar} for which \ref{ass:iid} - \ref{ass:stable} hold, and the innovation terms have finite fourth moments. Let $f\left(\cdot\right)$ be such that \ref{ass:fcond} holds and $\boldsymbol{\beta}^{f,j,\tau}$ exists. Using the definitions from \eqref{SVAR:eq:hat-def}, it then holds for $T \rightarrow \infty$
\begin{align*}
\hat{\beta}^{f,j,\tau}_k  & = \beta^{f,j,\tau}_k + {\scriptstyle \mathcal{O}}_p\left(1\right), \quad \widehat{\text{var}}\left(\hat{\beta}^{f,j,\tau}_k\right) = \mathcal{O}_p\left(\dfrac{1}{T}\right) \quad \text{and}\\
s_k^{j,\tau} & \overset{\mathbb{D}}{\to}\mathcal{N}\left(0,1\right) \ \forall k \not \in \text{AN}^{\tau}\left(j\right).
\end{align*}
\end{theorem}

\begin{algorithm}
\caption{Ancestor regression in SVAR}\label{alg:SVARAnc}
\begin{algorithmic}
\Inputs{$\mathbf{x} \in \mathbb R^{T \times d}$, degree $p$, target $j$, nonlinearity $f: \mathbb R \to \mathbb{R}$}

\State $\hat{\boldsymbol{\xi}}$ $\gets$ project out the contribution of $\mathbf{x}_{t-1}$ to $\mathbf{x}_{t-p}$ from $\mathbf{x}_t$

\For{$\tau = 0$ to $p$}
    \State $\hat{\boldsymbol{\xi}}^\tau$ $\gets$ project out the contribution of $\mathbf{x}_{t-\tau-1}$ to $\mathbf{x}_{t-\tau-p}$ from $\mathbf{x}_t$
    
    \State $\mathbf{s}^{j,\tau} \gets z$ values from OLS$\left(f\left(\hat{\boldsymbol{\xi}}_j^\tau\right) \sim \hat{\boldsymbol{\xi}}\right)$, where $\hat{\boldsymbol{\xi}}$ is appropriately lagged by $\tau$ steps
    \For{$k =1$ to $d$}
    \State $p_k^{j,\tau} = 2 \left\{1 - \Phi \left(\vert s_k^{j,\tau}\vert\right)\right\}$
    \EndFor
\EndFor
\Returns{$p_k^{j,\tau} \ \forall k,\tau$}
\end{algorithmic}
\end{algorithm}

\subsection{Simulation example}\label{sim-x4}
\begin{figure}[b!]
 \centering
 \includegraphics[width=0.8\textwidth]{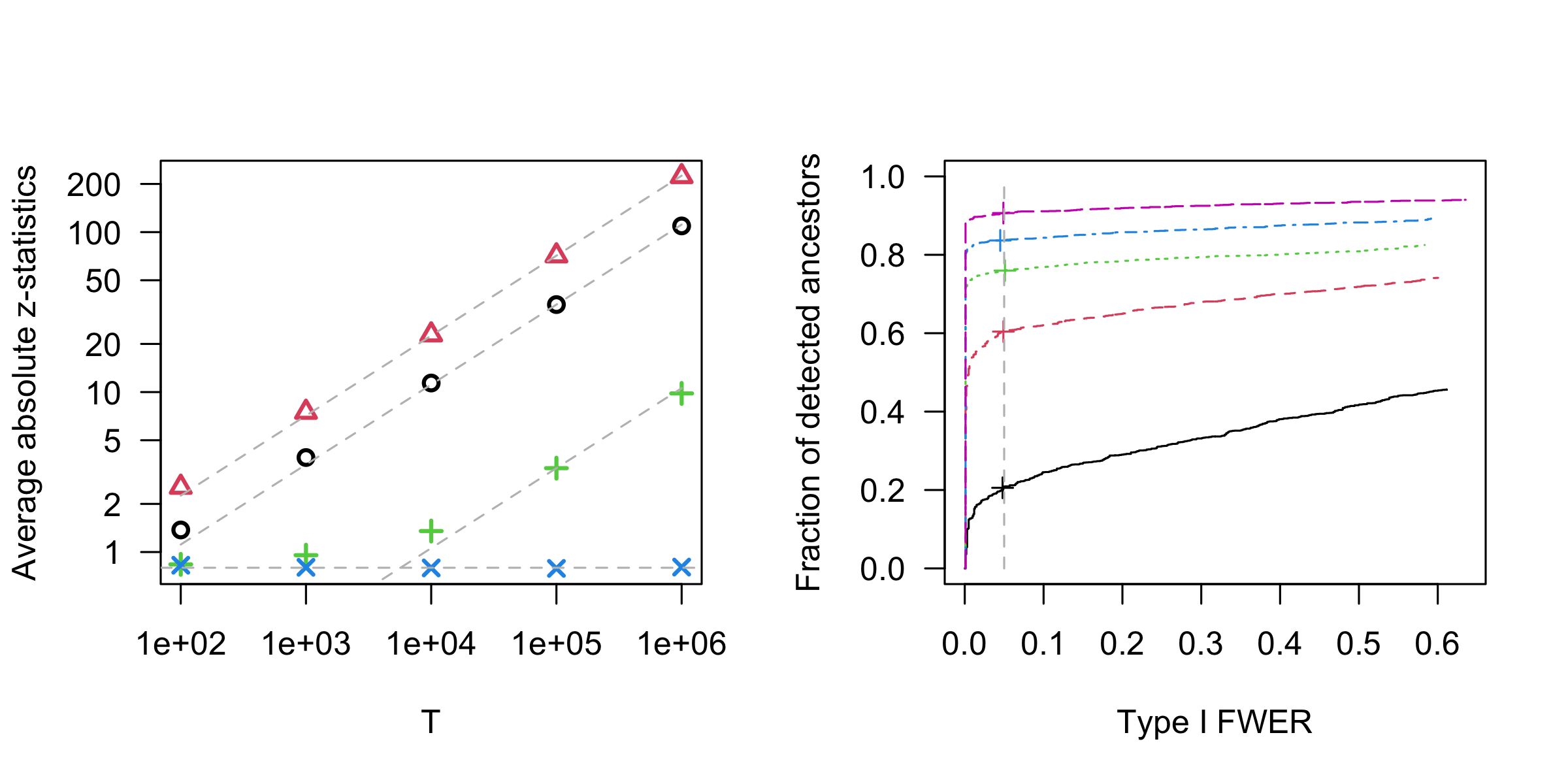}
 \caption[Simulation in a linear SEM]
 {Detecting the ancestors of $x_{t,4}$ in a structural vector autoregressive model of order $p=1$ with $6$ variables. The results are based on $1000$ simulation runs. On the left: Average absolute z-statistic for instantaneous ancestors (circles, black), lagged ancestors for which $\tilde{\mathbf{B}}_{4,k} \neq 0$ (triangles, red), lagged ancestors from which all causal paths start with an instantaneous effect (pluses, green), and non-ancestors (crosses, blue) for different sample sizes. The dashed diagonals correspond to $\surd{T}$-growth fitted to match at $T=10^5$ perfectly. The horizontal line corresponds to $\left(2/\pi\right)^{1/2}$, i.e., the first absolute moment of the asymptotic null distribution, a standard Gaussian. On the right: fraction of simulation runs with at least one false causal detection versus fraction of detected ancestors for the different sample sizes $10^2$ (solid, black), $10^3$ (dashed, red), $10^4$ (dotted, green), $10^5$ (dot-dashed, blue), and $10^6$ (long-dashed, pink). The curve uses the level $\alpha$ of the test as the implicit curve parameter. The pluses correspond to nominal $\alpha = 5\%$. The vertical line is at actual $5\%$.}
 \label{SVAR:fig:anc-x4-rand}
\end{figure}
We study ancestor regression in a small simulation example. We generate data from a structural vector autoregressive model with $d=6$ variables and order $p=1$. For the instantaneous effects, the causal order is fixed to be $x_{t,1}$ to $x_{t,6}$. Otherwise, the structure is randomized and changes per simulation run: $x_{t,k}$ is an instantaneous parent of $x_{t,l}$ for $k<l$ with probability $0.2$ such that there is an average of $3$ parental relationships. The edge weights are sampled uniformly and the distributions of the $\epsilon_{t,k}$ are assigned by permuting a fixed set of $6$ error distributions. The entries in $\mathbf{B}_1$ are non-zero with probability $0.1$. If so, they are sampled uniformly and assigned a random sign with equal probabilities. If the maximum absolute eigenvalue of $\tilde{\mathbf{B}}$ would be larger than $0.95$, $\mathbf{B}_1$ is shrunken such that this absolute eigenvalue is $0.95$ to ensure stability.

We aim to find the ancestors of $x_{t,4}$. We create $1000$ different setups and test each on sample sizes varying from $10^2$ to $10^6$. As a nonlinear function, we use $f\left(x_{t,j}\right) = x_{t,j}^3$. By z-statistic, we mean $s_k^{4,\tau}$ as in Theorem \ref{SVAR:theo:norm}. We calculate p-values according to \eqref{SVAR:eq:z-test} and apply a Bonferroni-Holm correction to them.

\begin{figure}[b!]
    \centering
\begin{tikzpicture}[ >=stealth, 
                    every node/.style={draw, circle, minimum size=1.5cm, text centered, font=\small}]

\node (x_t1_3) at (0, 0) {$x_{t-1,3}$};
\node (x_t1_4) [below=of x_t1_3] {$x_{t-1,4}$};

\node (x_t_3) [right=of x_t1_3] {$x_{t,3}$};
\node (x_t_4) [below=of x_t_3] {$x_{t,4}$};

\draw[->] (x_t1_4) -- (x_t_4);

\draw[->] (x_t_3) -- (x_t_4);

\draw[->] (x_t1_3) -- (x_t1_4);

\end{tikzpicture}
    \caption{Example of the different types of ancestors in Figure \ref{SVAR:fig:anc-x4-rand}.}\label{fig:illu}
\end{figure}
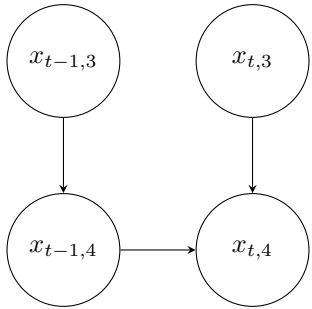

On the left-hand side of Figure \ref{SVAR:fig:anc-x4-rand}, we see the average absolute z-statistics for ancestors and non-ancestors for the different sample sizes. We distinguish between three types of ancestors: instantaneous ancestors, lagged ancestors for which $\tilde{\mathbf{B}}_{4,k} \neq 0$, and lagged ancestors from which all causal paths start with an instantaneous effect. Figure \ref{fig:illu} illustrates the different types of lagged ancestors. The path from $x_{t-1,4}$ to $x_{t,4}$ is not mediated by instantaneous effects at $t-1$ and $\left(\tilde{\mathbf{B}}_1\right)_{4,4} \neq 0$.
Such lagged ancestors have stronger signals than instantaneous ones. This agrees with the intuition that it is easier to find a directed causal path if it is a priori known that only one direction could be possible, i.e., the lagged variable can be an ancestor but not a descendant such that one type of dependence is excluded. However, the only causal path from $x_{t-1,3}$ to $x_{t,4}$ starts with an instantaneous effect such that $\left(\tilde{\mathbf{B}}_1\right)_{4,3} = 0$. To detect such an effect, one must essentially detect that instantaneous effect but with a target that contains additional noise from all other contributions. Hence, this is harder than detecting instantaneous effects. As we randomly sample the edge weights from a continuous distribution, there are no unfaithful zeros in $\tilde{\mathbf{B}}$.

For non-ancestors, the observed average of the absolute z-statistics is close to the theoretical mean under the asymptotic null distribution as desired. On the right-hand side, we see that we can control the type I error at the desired level for every sample size. As expected, the power to detect ancestors increases with larger sample sizes. However, driven by the last group of ancestors discussed above, there are still some undetected ancestors for $T=10^6$. For the other groups, we obtain almost perfect power. One could then also infer these missed effects by recursive arguments. We discuss this for the case of networks below.

\section{Inferring effects in networks}
So far, we assumed that there is a time series component $x_{t,j}$ whose causal ancestors are of special interest. This is not always the case. Instead, one might be interested in inferring the full set of causal connections between the variables. Naturally, our ancestor detection technique can be extended to that problem by applying it nodewise. After estimating the effects on every time series, there is a total of $\left(p+1\right)d\left(d-1\right)$ p-values to consider when ignoring autoregressive effects.
We suggest the construction of two types of ancestral graphs that could be of interest.

\subsection{Instantaneous effects}\label{inst-graph}
\begin{algorithm}[b!]
\caption{Instantaneous graph}\label{alg:inst}
\begin{algorithmic}
\Inputs{$\mathbf{x} \in \mathbb R^{T \times d}$, degree $p$, nonlinearity $f: \mathbb R \to \mathbb{R}$, level $\alpha$}

\For{$j = 1$ to $d$}
\State Use Algorithm \ref{alg:SVARAnc} with target $j$ to get $p_k^{j,0} \ \forall k\neq j$
\EndFor
\State Apply Bonferroni-Holm correction over all $d\left(d-1\right)$ tests to obtain multiplicity corrected $P_k^{j,0}$
\State Initiate $\widehat{\text{AN}}^{0}\left(j\right) = \emptyset \forall j$

\For{$j,k = 0$ to $d$}

\If{$P_k^{j,0} < \alpha$}
\State $\widehat{\text{AN}}^{0}\left(j\right) = \left\{\widehat{\text{AN}}^{0}\left(j\right),k\right\}$ \Comment{Find initial estimated set using significant tests}
\EndIf
\EndFor

\While{$\exists j,k,l$ such that $k \in \widehat{\text{AN}}^{0}\left(j\right)$, $l \in \widehat{\text{AN}}^{0}\left(k\right)$ but $l \notin \widehat{\text{AN}}^{0}\left(j\right)$}
\State \State $\widehat{\text{AN}}^{0}\left(j\right) = \left\{\widehat{\text{AN}}^{0}\left(j\right),l\right\}$ \Comment{Add ancestors recursively}
\EndWhile

\State Initiate $\tilde{\alpha} = \alpha$
\While{$\exists j$ such that $j \in \widehat{\text{AN}}^{0}\left(j\right)$}
\State $I = \left\{j: \quad j \in \widehat{\text{AN}}^{0}\left(j\right)\right\}$ \Comment{Variables that are on a cycle}
\State Find highest p-value $\tilde{\alpha} \coloneqq P_l^{k,0} < \alpha$, where $k,l \in I$
\State Re-estimate ancestor relations within I using level $\tilde{\alpha}$
\EndWhile
\Returns{$\widehat{\text{AN}}^{0}\left(j\right) \quad \forall j$ and $\tilde{\alpha}$}
\end{algorithmic}
\end{algorithm}
Focusing on instantaneous effects only, the situation is very similar as in the i.i.d.\ case discussed in \cite{schultheiss2023ancestor}. Hence, we apply the same algorithm:

First, we apply a multiplicity correction over the $d\left(d-1\right)$ tests to control the type I family-wise error rate. We use the Bonferroni-Holm multiplicity correction.
Then, we construct further ancestral relationships recursively: E.g., if $x_{t,1}$ has an instantaneous effect on $x_{t,2}$, and $x_{t,2}$ has an instantaneous effect on $x_{t,3}$, there must be an instantaneous effect from $x_{t,1}$ to $x_{t,3}$. If all detected effects are correct, all such recursively constructed effects must be correct as well. Hence, the type I family-wise error rate remains the same while the power can increase and typically does for larger networks.

\begin{figure}
\begin{tikzpicture}[scale=0.8, every node/.style={scale=0.8}]
  \node[circle, draw] (x1) at (0,0) {$x_{t,1}$};
  \node[circle, draw] (x2) at (3,0) {$x_{t,2}$};
  \node[circle, draw] (x4) at (3,-3) {$x_{t,4}$};
  \node[circle, draw] (x3) at (0, -3) {$x_{t,3}$};

  \draw[->] (x1) -- (x2) node[midway, above, font=\small] {$10^{-3}$};
  \draw[->] (x2) -- (x4) node[midway, right, font=\small] {$10^{-2}$};
  \draw[->] (x3) -- (x1) node[midway, left, font=\small] {$10^{-2}$};
  \draw[->] (x2) -- (x3) node[midway, right, font=\small] {$10^{-4}$};

  \node[circle, draw] (x1b) at (5,0) {$x_{t,1}$};
  \node[circle, draw] (x2b) at (8,0) {$x_{t,2}$};
  \node[circle, draw] (x4b) at (8,-3) {$x_{t,4}$};
  \node[circle, draw] (x3b) at (5, -3) {$x_{t,3}$};

  \draw[<->] (x1b) -- (x2b) ;
  \draw[->] (x1b) -- (x4b) ;
  \draw[->] (x2b) -- (x4b) ;
  \draw[->] (x3b) -- (x4b) ;
  \draw[<->] (x3b) -- (x1b) ;
  \draw[<->] (x2b) -- (x3b) ;

  \node[circle, draw] (x1c) at (10,0) {$x_{t,1}$};
  \node[circle, draw] (x2c) at (13,0) {$x_{t,2}$};
  \node[circle, draw] (x4c) at (13,-3) {$x_{t,4}$};
  \node[circle, draw] (x3c) at (10, -3) {$x_{t,3}$};

  \draw[->] (x1c) -- (x2c) node[midway, above, font=\small] {$10^{-3}$};
  \draw[->] (x2c) -- (x4c) node[midway, right, font=\small] {$10^{-2}$};
  \draw[->] (x2c) -- (x3c) node[midway, right, font=\small] {$10^{-4}$};

  \node[circle, draw] (x1d) at (15,0) {$x_{t,1}$};
  \node[circle, draw] (x2d) at (18,0) {$x_{t,2}$};
  \node[circle, draw] (x4d) at (18,-3) {$x_{t,4}$};
  \node[circle, draw] (x3d) at (15, -3) {$x_{t,3}$};

  \draw[->] (x1d) -- (x2d) ;
  \draw[->] (x1d) -- (x4d) ;
  \draw[->] (x2d) -- (x4d) ;
  \draw[->] (x1d) -- (x3d) ;
  \draw[->] (x2d) -- (x3d) ;
\end{tikzpicture}
    \caption{From left to right: Detected edges with $\alpha = 0.05$ and the corresponding p-values. Recursive construction leading to cycles. Detected edges with $\alpha < 0.01$ and the corresponding p-values. Recursive construction without cycles.}\label{fig:inst}
\end{figure}
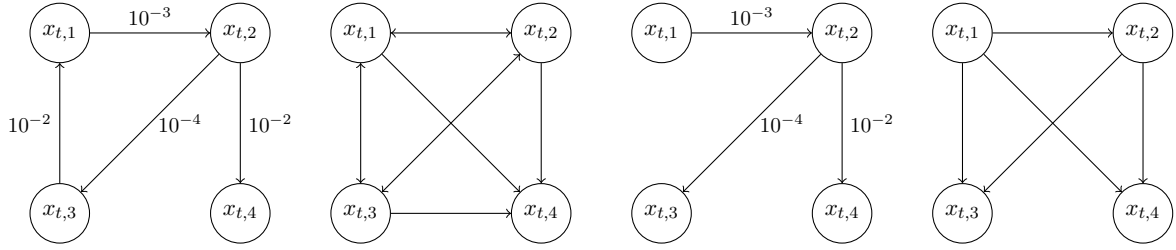

If we make type I errors, this could create contradictions leading to cycles.
Then, we gradually decrease the significance level for edges within these cycles until no more cycles remain. The largest significance level for which no loops occur is also an asymptotically valid p-value for the null hypothesis that the data come from model \eqref{eq:var} with assumptions \ref{ass:iid} - \ref{ass:fcond} as we have asymptotic error control under this model class. An example where $\alpha = 0.05$ leads to cycles is demonstrated in Figure \ref{fig:inst}. We keep the edge from $x_{t,2}$ to $x_{t,4}$ using the initial significance level as it is not part of any cycles. The procedure is summarized in Algorithm \ref{alg:inst}.

\subsection{Summary time graph}\label{sum-graph}
\begin{algorithm}[b!]
\caption{Summary graph}\label{alg:summ}
\begin{algorithmic}
\Inputs{$\mathbf{x} \in \mathbb R^{T \times d}$, degree $p$, nonlinearity $f: \mathbb R \to \mathbb{R}$, level $\alpha$}

\For{$j = 1$ to $d$}
\State Use Algorithm \ref{alg:SVARAnc} with target $j$ to get $p_k^{j,\tau} \ \forall k\neq j, \tau \leq p$
\For{$k \neq j$}
\State Obtain $p_k^{j}$ from $p_k^{j,0}, \ldots, p_k^{j,p}$ as in Appendix \ref{app:comb}
\EndFor
\EndFor
\State Apply Bonferroni-Holm correction over all $d\left(d-1\right)$ tests to obtain multiplicity corrected $P_k^{j}$
\State Initiate $\widehat{\text{AN}}\left(j\right) = \emptyset \forall j$

\For{$j,k = 0$ to $d$}
\If{$P_k^{j,0} < \alpha$}
\State $\widehat{\text{AN}}\left(j\right) = \left\{\widehat{\text{AN}}\left(j\right),k\right\}$ \Comment{Find initial estimated set using significant tests}
\EndIf
\EndFor

\While{$\exists j,k,l$ such that $k \in \widehat{\text{AN}}^{0}\left(j\right)$, $l \in \widehat{\text{AN}}^{0}\left(k\right)$ but $l \notin \widehat{\text{AN}}\left(j\right)$}
\State \State $\widehat{\text{AN}}\left(j\right) = \left\{\widehat{\text{AN}}\left(j\right),l\right\}$ \Comment{Add ancestors recursively}
\EndWhile

\Returns{$\widehat{\text{AN}}\left(j\right) \quad \forall j$}
\end{algorithmic}
\end{algorithm}
If not only instantaneous effects but all effects are of interest, one can consider a summary time graph. It includes an edge from $k$ to $j$ if there is a causal path from $x_{t',k}$ to $x_{t,j}$ for any $t' \leq t$. This graph can be cyclic under our modeling assumptions.

To obtain it, we first assign a p-value to each potential edge $k \rightarrow j$, say, $p_k^j$. There are $p+1$ p-values corresponding to this edge, i.e., $p_k^{j,0}, \ldots p_k^{j,p}$. We combine them using ideas from \cite{meinshausen2009p} designed to combine p-values under arbitrary dependence. For details, see Appendix \ref{app:comb}. Again, we apply the Bonferroni-Holm multiplicity correction to control the family-wise error rate. This allows to add recursively more edges while still controlling the type I error rate. Model \eqref{eq:svar} does not imply that the summary graph must be acyclic. Thus, we output the result after the recursive construction even in the presence of cycles. For example, in Figure \ref{fig:inst} if the depicted p-values are (multiplicity corrected) summary p-values, the obtained summary graph is the second to the left. We summarize the procedure in Algorithm \ref{alg:summ}.

In Figure \ref{SVAR:fig:anc-x4-rand}, we saw that lagged ancestors from which the directed path begins with an instantaneous edge are the hardest to detect. Here, such a recursive construction can help. Assume $x_{t,1} \rightarrow x_{t,2} \rightarrow x_{t+1,3}$. Then, there is also a detectable effect from $x_{t,1}$ to $x_{t+1,3}$, but it can be easier to detect the two intermediate edges.

\subsection{Simulation example}\label{network-sim}
\begin{figure}[b!]
 \centering
 \includegraphics[width=0.8\textwidth]{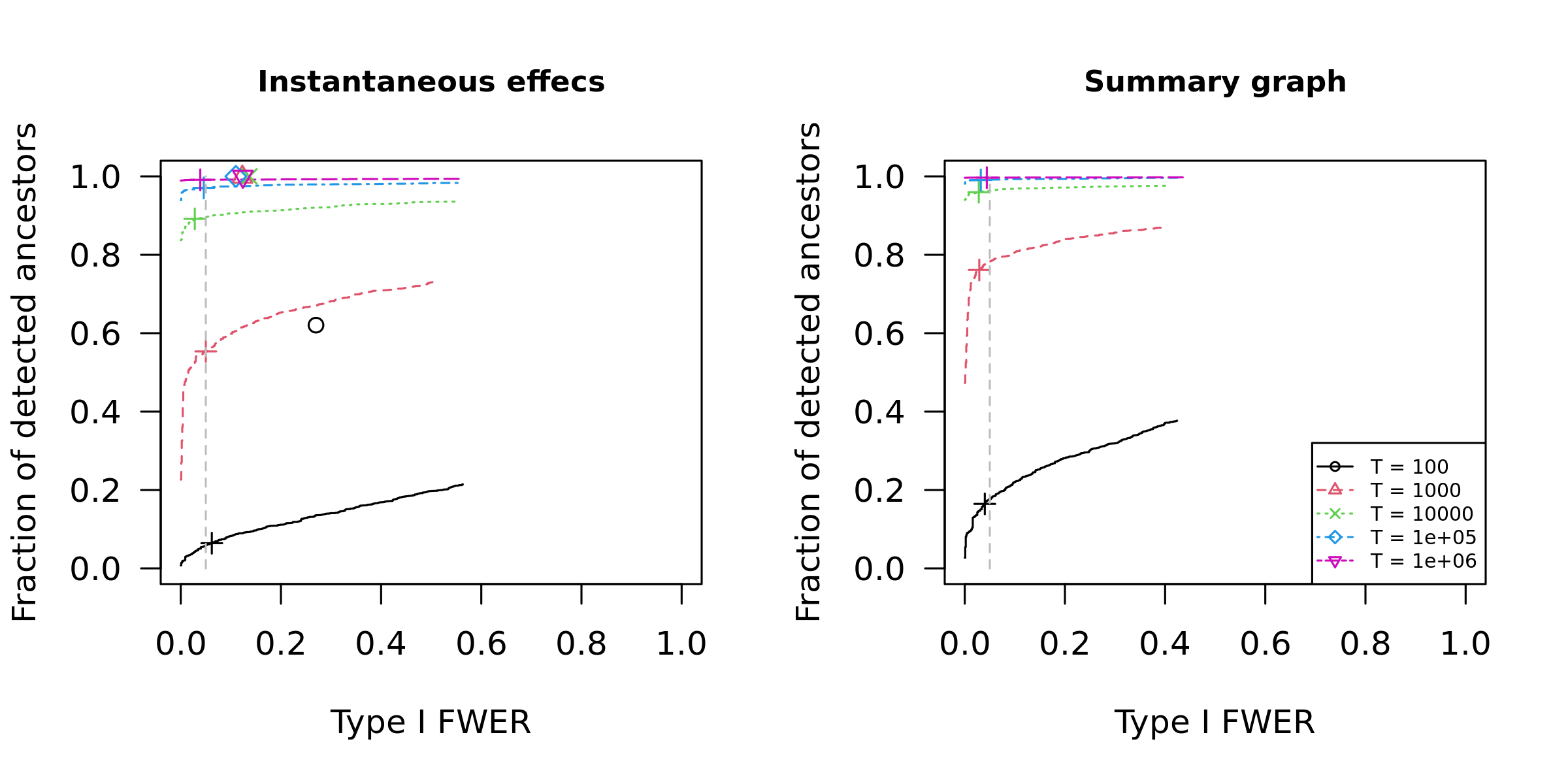}
 \caption[Simulation in a linear SEM]
 {Nodewise ancestor detection in a structural vector autoregressive model of order $p=1$ with $6$ variables. The results are based on $1000$ simulation runs. Depicted is the family-wise error rate of false causal detection versus the fraction of detected ancestors. The curves use the level of the test $\alpha$ as an implicit curve parameter. The pluses correspond to nominal $\alpha = 5\%$. The vertical line is at actual $5\%$. The other symbols represent the performance of the LiNGAM algorithm. On the left: instantaneous effects. On the right: summary graph.}
 \label{SVAR:fig:graph-perf-rand}
\end{figure}
We extend the simulation in \ref{sim-x4} to the network setting. We estimate both, an instantaneous ancestral graph as in Section \ref{inst-graph} and a summary time graph as in Section \ref{sum-graph}.

In Figure \ref{SVAR:fig:graph-perf-rand}, we show the obtained detection rate versus the type I family-wise error rate for varying significance levels. For the summary graph, we obtain slightly better performance. This matches the intuition that this less detailed information is easier to obtain. For either, we achieve essentially perfect separation between ancestors and non-ancestors for large enough sample sizes. Hence, the recursive construction helped to detect even these effects that appeared hard to find based on the individual test as in Figure \ref{SVAR:fig:anc-x4-rand}. For the instantaneous effects, there is a slight overshoot of the type I error for $T=100$, i.e., the asymptotic null distribution is not sufficiently attained yet. For all longer time series, it is controlled as desired.

Additionally, we show on the left side the performance of the LiNGAM algorithm as described in \cite{hyvarinen2010estimation}. For this, we use the code published together with \cite{moneta2013causal}. As the LiNGAM algorithm by default does not search for sparse estimates $\hat{\mathbf{B}}_\tau$ for $\tau>0$, a straightforward comparison is only possible for instantaneous effects. We note that the LiNGAM algorithm leads to higher power, especially for low sample sizes. However, it does not allow for an interpretable error control at a predefined level. Similar results were obtained for the comparison in the i.i.d.\ case, see \cite{schultheiss2023ancestor}.

\begin{figure}[b!]
 \centering
 \includegraphics[width=0.8\textwidth]{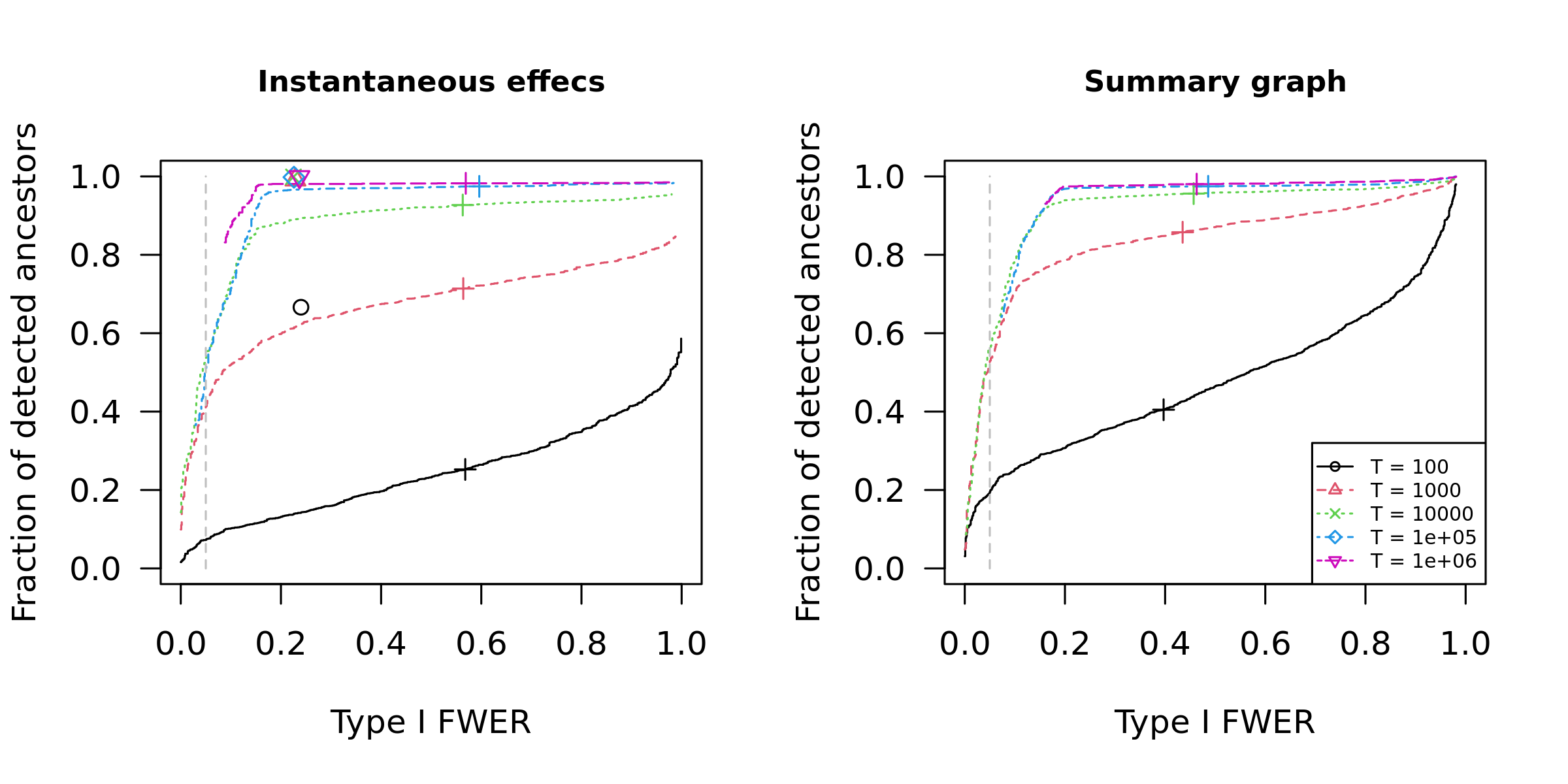}
 \caption[Simulation in a linear SEM]
 {Nodewise ancestor detection in a structural vector autoregressive model of order $p=1$ with $6$ variables, whereof one is unobserved. See the caption of Figure \ref{SVAR:fig:graph-perf-rand} for explanations.}
 \label{SVAR:fig:graph-perf-rand-hid}
\end{figure}

In the next experiment, we analyze the effect of unobserved variables. We reuse the same simulation setting but randomly consider one of the time series as unobserved for each simulation run. We run ancestor regression and LiNGAM with the observed time series and calculate the same performance statistics. The results are in Figure \ref{SVAR:fig:graph-perf-rand-hid}.

Control of the familywise error rate at a fixed level no longer works for ancestor regression. However, there is still a separation between ancestors and non-ancestors in terms of the effect size. For long time series, there is a sweet spot with high power at a fairly low error rate. Hence, the ordering of the p-values still gives some indication of what could be the true ancestors. Some of the curves do not start at $0$ power and error rate as there are p-values that are numerically equal to $0$ such that no reasonable ordering can be made for these observations. 

At our target level $\alpha=0.05$, the power remains comparable to the case without hidden variables or is even increased for some sample sizes. The unobserved variables are mainly a problem for error control as the assumptions are not fulfilled but not for detection per se. Similarly, the power of the LiNGAM algorithm remains high while the error rate is increased.

Additional simulation results with higher numbers of time series, both with and without unobserved variables are in Appendix \ref{app:high}.

\section{Real data applications}\label{real-data}
Inspired by \cite{peters2013causal}, we apply our method to several bivariate time series as proof of concept. As they suggest, we fit models of order $p=6$.

\subsection*{Old Faithful geyser}
We analyze data from the Old Faithful geyser \citep{azzalini1990look} provided in the \textsf{R}-package \texttt{MASS} \citep{venables2002modern}. It contains information on the waiting time leading to an eruption ($x_{t,1}$) and the duration of an eruption ($x_{t,2}$) for $299$ consecutive eruptions. We model these as a bivariate time series although we do not have the classical framework with equidistant measurements in time. The consensus is that the eruption duration affects the subsequent waiting time more than vice-versa. 

Our algorithm outputs no significant instantaneous effects on this dataset ($p_2^{1,\tau =0} = 0.78$, $p_1^{2,\tau =0} = 0.73$). This is in line with the consensus which suggests no instantaneous effect from waiting to duration. Here, the duration corresponds to something that happened after the waiting of the corresponding time point such that there should neither be an ``instantaneous'' effect from duration to waiting. Our summarized p-values suggest an effect from duration to waiting ($p_2^1 = 15 * 10^{-22}$) while the opposite direction is borderline significant ($p_1^2 = 0.094$).
\begin{figure}[h!]
    \centering
\begin{tikzpicture}
    \node[draw, circle, minimum size=1.8cm] (waiting) at (0, 0) {waiting};
    \node[draw, circle, minimum size=1.8cm] (duration) at (4, 0) {duration};

    \draw[dotted, ->] (waiting) to[bend left] node[above] {$p_1^{2,\tau = 0}=0.73$} (duration);
    \draw[dotted, ->] (duration) to[bend left] node[below] {$p_2^{1, \tau = 0} = 0.78$} (waiting);

    \node[draw, circle, minimum size=1.8cm] (waiting2) at (8, 0) {waiting};
    \node[draw, circle, minimum size=1.8cm] (duration2) at (12, 0) {duration};

    \draw[dotted, ->] (waiting2) to[bend left] node[above] {$p_1^2=0.094$} (duration2);
    \draw[->] (duration2) to[bend left] node[below] {$p_2^{1} = 15 * 10^{-22}$} (waiting2);
\end{tikzpicture}
    \caption{Estimated effects for the geyser data. On the left, p-values corresponding to instantaneous effects. On the right, summarized p-values over all considered lags, see Section \ref{sum-graph}. Significant edges are drawn as full lines, the others are dotted.}
\end{figure}
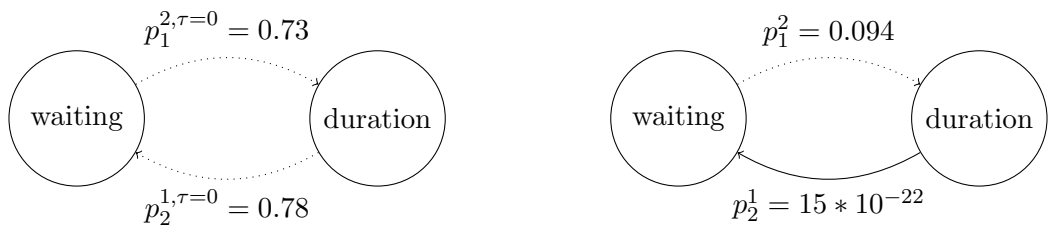

\begin{figure}[b!]
    \centering
\begin{tikzpicture}
    \node[draw, circle, minimum size=1.8cm] (waiting) at (0, 0) {waiting};
    \node[draw, circle, minimum size=1.8cm] (duration) at (4, 0) {duration};

    \draw[dotted, ->] (waiting) to[bend left] node[above] {$p_1^{2,\tau =0} = 0.51$} (duration);
    \draw[->] (duration) to[bend left] node[below] {$p_2^{1,\tau =0} = 5*10^{-4}$} (waiting);

    \node[draw, circle, minimum size=1.8cm] (waiting2) at (8, 0) {waiting};
    \node[draw, circle, minimum size=1.8cm] (duration2) at (12, 0) {duration};

    \draw[dotted, ->] (waiting2) to[bend left] node[above] {$p_1^2 =0.18$} (duration2);
    \draw[->] (duration2) to[bend left] node[below] {$p_2^1 = 9*10^{-3}$} (waiting2);
\end{tikzpicture}
    \caption{Estimated effects for the shifted geyser data. On the left, p-values corresponding to instantaneous effects. On the right, summarized p-values over all considered lags, see Section \ref{sum-graph}. Significant edges are drawn as full lines, the others are dotted.}
\end{figure}
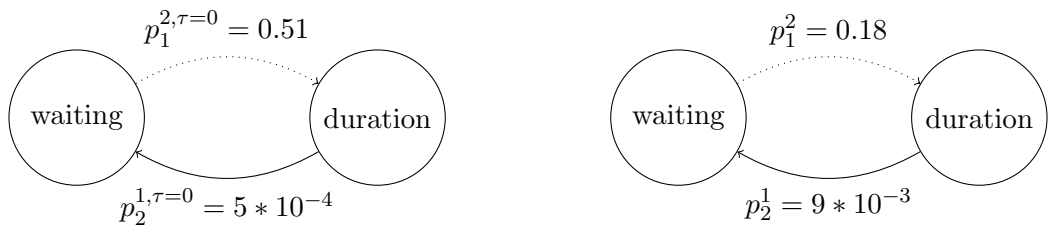
We consider a slightly altered time series shifted such that waiting corresponds to the waiting after the given eruption, i.e., $298$ observations remain. Now, our output suggests an instantaneous effect from duration to waiting ($p_2^{1,\tau =0} = 5*10^{-4}$, $p_1^{2,\tau =0} = 0.51$) in agreement with the consensus belief. The summarized p-values are $p_2^1 = 9*10^{-3}$ and $p_1^2 =0.18$ respectively.

\subsection*{Gas furnace}
We look at data from a gas furnace described in \cite{box2015time}. It can be downloaded from \href{https://openmv.net/info/gas-furnace}{https://openmv.net/info/gas-furnace}. The time series are the input gas rate ($x_{t,1}$) and the output $CO_2$ level observed at $296$ equidistant time points. The more plausible causal direction is from input to output.

Our algorithm outputs no instantaneous effects ($p_1^{2,\tau =0} = 0.18$, $p_2^{1,\tau =0} = 0.55$). But, over lags, there appears to be an effect from the input rate to the output concentration as expected. ($p_1^2 = 4 * 10^{-20}$, $p_2^1 = 1$).
\begin{figure}[h]
    \centering
\begin{tikzpicture}
    \node[draw, circle, minimum size=1.8cm] (gasrate) at (0, 0) {gas rate};
    \node[draw, circle, minimum size=1.8cm] (co2) at (4, 0) {$CO_2$};

    \draw[dotted, ->] (gasrate) to[bend left] node[above] {$p_1^{2,\tau =0} = 0.18$} (co2);
    \draw[dotted, ->] (co2) to[bend left] node[below] {$p_2^{1,\tau =0} = 0.55$} (gasrate);

    \node[draw, circle, minimum size=1.8cm] (gasrate2) at (8, 0) {gas rate};
    \node[draw, circle, minimum size=1.8cm] (co22) at (12, 0) {$CO_2$};

    \draw[->] (gasrate2) to[bend left] node[above] {$p_1^2 = 4 * 10^{-20}$} (co22);
    \draw[dotted, ->] (co22) to[bend left] node[below] {$p_2^1 = 1$} (gasrate2);
\end{tikzpicture}
    \caption{Estimated effects for the gas furnace data. On the left, p-values corresponding to instantaneous effects. On the right, summarized p-values over all considered lags, see Section \ref{sum-graph}. Significant edges are drawn as full lines, the others are dotted.}
\end{figure}
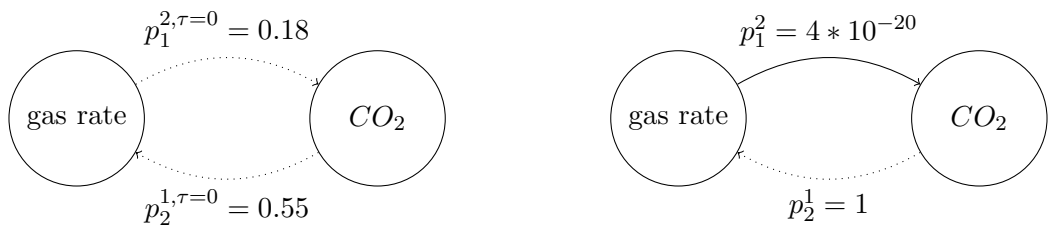

\subsection*{Dairy}
We use data on ten years of weekly prices for butter ($x_{t,1}$) and cheddar cheese ($x_{t,2}$), i.e., $522$ observations in total. \cite{peters2013causal} present this as an example where the price of milk could act as a hidden confounder hence violating the model assumptions. Unfortunately, the data source that they mention has disappeared. But, the data was kindly provided by the first author.

We detect no significant instantaneous effects ($p_1^{2,\tau =0} = 0.64$, $p_2^{1,\tau =0} = 0.71$) and hence also no model violations. There is a significant lagged effect from butter to cheddar ($p_1^2 = 5 * 10^{-15}$, $p_2^1 = 1$). In the case of a hidden confounder, both effects should appear in the summary time graph, but we have no evidence for this. However, given the size of the dataset, it can well be that we missed the spurious effect from cheddar to butter.
\begin{figure}[h]
    \centering
\begin{tikzpicture}
    \node[draw, circle, minimum size=1.8cm] (butter) at (0, 0) {butter};
    \node[draw, circle, minimum size=1.8cm] (cheddar) at (4, 0) {cheddar};

    \draw[dotted, ->] (butter) to[bend left] node[above] {$p_1^{2,\tau =0} = 0.64$} (cheddar);
    \draw[dotted, ->] (cheddar) to[bend left] node[below] {$p_2^{1,\tau =0} = 0.71$} (butter);

    \node[draw, circle, minimum size=1.8cm] (butter2) at (8, 0) {butter};
    \node[draw, circle, minimum size=1.8cm] (cheddar2) at (12, 0) {cheddar};

    \draw[->] (butter2) to[bend left] node[above] {$p_1^2 = 5 * 10^{-15}$} (cheddar2);
    \draw[dotted, ->] (cheddar2) to[bend left] node[below] {$p_2^1 = 1$} (butter2);
\end{tikzpicture}
    \caption{Estimated effects for the dairy data. On the left, p-values corresponding to instantaneous effects. On the right, summarized p-values over all considered lags, see Section \ref{sum-graph}. Significant edges are drawn as full lines, the others are dotted.}
\end{figure}
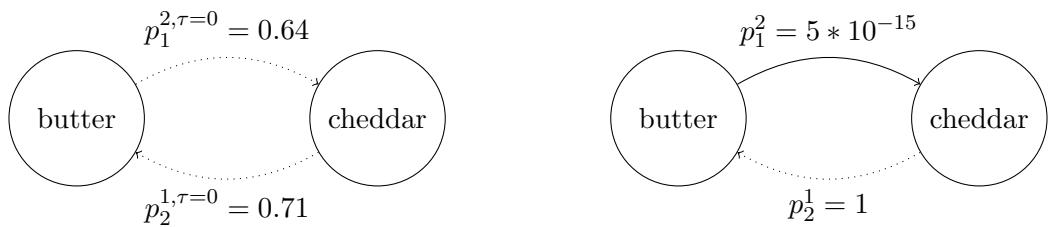

\section{Discussion}
\subsection{Outlook: Lessons for independent data with background knowledge}
Regressing $\xi_{t,j}^{\tau}$ against $\boldsymbol{\xi}_{t-\tau}$ instead of $x_{t,j}$ against $\mathbf{x}_{t-\tau}$ means that we first project out all other covariates that might have a confounding effect but are surely not descendants. Thus, all effects from time points before $t-\tau$ are taken out of the analysis, and the fraction of relevant information in the data increases. If we projected out only after the transformation, i.e., use $f\left(x_{t,j}\right)$ those effects could not be fully taken out, and the noise level increases. 

Similarly, it can happen that for data with no time structure a certain variable is known to be a (potential) confounder between others, but surely not a descendant of any of these. For example, if we measure the weight and height of children, our common sense says that age is a confounder of the two but surely not causally affected by either. In our assumed framework, i.e., linear causal relations, one can then first regress out this confounding effect which can be done perfectly (in population) before applying the transformation. If it is done after the transformation, there remains some noise terms stemming from the confounder which decreases the signal-to-noise ratio for the effects we are truly interested in.

Importantly, not every variable that is of lesser interest can be regressed out a priori. If its relative place in the causal order is not known, it has to be included in the usual way to retain type I error guarantees. Of course, one can always omit the according tests corresponding to uninteresting effects such that less multiplicity correction must be applied.

\subsection{Conclusion}
We introduce a new method for causal discovery in structural vector autoregressive models. We assess whether there is a causal effect from one time series component to another for any given time lag. The method is computationally very efficient and has asymptotic type I error control against false causal discoveries. Our simulations show that this error control works well for finite time series as well.

We also obtain asymptotic power up to few pathological cases. In networks, additional effects can be inferred by the logic that an ancestor of an ancestor must be an ancestor. In our simulation, we see that this can help to find almost all ancestors without errors even when some connections are individually hard to find at a given sample size. When incorporating unobserved time series, which is a violation of our assumptions, error control at a predefined level does not work anymore. However, the ordering of the test statistics still provides some indication of what could be true ancestors.

We apply our method to three real-world bivariate time series and obtain results that mostly agree with the common understanding of the underlying process. Hence, we demonstrate that ancestor regression can be of use even when the modeling assumptions are not fulfilled as in the ideal simulated cases, and the data is only of medium size.

Code scripts to reproduce the results presented in this paper are available here\\ \textbf{\href{https://github.com/cschultheiss/SVAR-Ancestor}{https://github.com/cschultheiss/SVAR-Ancestor}}.\\

\noindent \textbf{Acknowledgment:} The project leading to this application has received funding from the European Research Council (ERC) under the European Union’s Horizon 2020 research and innovation programme (grant agreement No 786461).\\
 
\bibliographystyle{apalike} 
\bibliography{references}

\newpage
\appendix
\allowdisplaybreaks

\newpage
\section{Proofs}\label{SVAR:app:proof}
\subsection{Additional notation}
We introduce additional notation that is used for the proofs. We do not explicitly mention the time steps considered for a regression estimate. It is always meant to use as many observations as available.
The number of observations used for an estimate we call simply $T$ as $T \rightarrow \infty$ is equivalent to $T - p - \tau \rightarrow \infty$.

Subindexing a matrix or vector containing several time lags e.g., $\mathbf{x}_{\_p+1,k}$ means only selecting the column or entry corresponding to time series $k$ with no time lag unless stated otherwise. Subindex $-k$ means all but this column or entry.
$I_T$ is the $T$-dimensional identity matrix.
$P_{-k}$ denotes the orthogonal projection onto $\mathbf{x}_{\_p+1,-k}$ and $P^{\perp}_{-k} = I_T - P_{-k}$ denotes the orthogonal projection onto its complement. $P_{\mathbf{x}_{-\tau}}$ is the orthogonal projection onto all $\mathbf{x}_{-\tau-1\_p}$. 

For some random vector $\mathbf{x}_{t\_p+1}$, we have the moment matrix $\boldsymbol{\Sigma^{\mathbf{x}}} \coloneqq \EE\left[\mathbf{x}_{t\_p+1}\mathbf{x}_{t\_p+1}^\top\right]$. This equals the covariance matrix for centered $\mathbf{x}_{t\_p+1}$. We assume this matrix to be invertible. Then, the principal submatrix $\boldsymbol{\Sigma}^\mathbf{x}_{-j,-j} \coloneqq \EE\left[\mathbf{x}_{t\_p+1,-j}\mathbf{x}_{t\_p+1,-j}^\top\right]$ is also invertible. Again, the negative subindex means the realization without time lag is omitted. We make the analogous assumption for $\boldsymbol{\Sigma^{\boldsymbol{\xi}}} \coloneqq \EE\left[\boldsymbol{\xi}_{t}\boldsymbol{\xi}_{t}^\top\right]$.

\subsection{Previous work}
We adapt some definitions from and results proved in \cite{schultheiss2023higher}.
\begin{equation} \label{eq:z-w-def}
 \begin{aligned}
z_{t,k} & \coloneqq x_{t,k} - \mathbf{x}_{t\_p+1,-k}^{\top}\boldsymbol{\gamma}_k, \quad &&\text{where}\\
\boldsymbol{\gamma}_k & \coloneqq \underset{\mathbf{b} \in \mathbb{R}^{d\left(p+1\right)-1}}{\text{argmin}}\EE\left[\left(x_{t,k} - \mathbf{x}_{t\_p+1,-k}^{\top} \mathbf{b}\right)^2\right] = \left(\Sigma^\mathbf{x}_{-k,-k}\right)^{-1}\EE\left[\mathbf{x}_{t\_p+1,-k}x_{t,k}\right],\\
w_{t,k} & \coloneqq f\left(\xi_{t+\tau,j}^{\tau}\right) - \boldsymbol{\xi}_{t,-k}^{\top}\boldsymbol{\zeta}_k, \quad &&\text{where}\\
 \boldsymbol{\zeta}_k & \coloneqq \underset{\mathbf{b} \in \mathbb{R}^{d-1}}{\text{argmin}}\EE\left[\left(f\left(\xi_{t+\tau,j}^{\tau}\right) - \boldsymbol{\xi}_{t,-k}^{\top} \mathbf{b}\right)^2\right] = \left(\Sigma^{\boldsymbol{\xi}}_{-k,-k}\right)^{-1}\EE\left[\boldsymbol{\xi}_{t,-k}f\left(\xi_{t+\tau,j}^{\tau}\right)\right].
 \end{aligned}
\end{equation}
We denote by $\mathbf{z}_k \coloneqq \mathbf{x}_{\_p+1,k} - \mathbf{x}_{\_p+1,-k} \boldsymbol{\gamma}_k$ and $\mathbf{w}_k$ analogously these true regression residuals at the relevant time points. For notational simplicity, we do not index $\mathbf{w}_k$ and $\boldsymbol{\zeta}_k$ with the lag $\tau$ as the arguments remain the same for every fixed lag.
Using these definitions, we have
\begin{align*}
\beta^{f,j,\tau}_k = \EE\left[z_{t,k} w_{t,k}\right]/\EE\left[z^2_{t,k}\right]=\EE\left[z_{t,k} f\left(\xi_{t,j}^{\tau}\right)\right]/\EE\left[z^2_{t,k}\right].
\end{align*}
from partial regression.
\subsection{Proof of Theorem \ref{theo:bols}}\label{app:proof:bols}
Under \eqref{eq:svar}, $\mathbf{x}_{t\_p+1}$ includes all causal parents of $x_{t,k}$. Now an argument analogous to Lemma 5 in the supplemental material of \cite{schultheiss2023higher} shows that $z_{t,k}$ must be a linear combination of $\epsilon_{t,k}$ and possibly some $\epsilon_{t,l}$ where $k \in \text{AN}^0\left(l\right)$. If $k \not \in \text{AN}^\tau\left(j\right)$, $x_{t+\tau,j}$ must be independent of these innovation terms. Furthermore, $z_{t,k} \perp \mathbf{x}_{t'} \ \forall t' <t$. Hence,
\begin{equation*}
z_{t,k} \perp \xi^{\tau}_{t+\tau,j} = x_{t+\tau,j} - \left(\mathbf{A}_{\tau}\right)_j \mathbf{x}_{t-1\_p},
\end{equation*}
using \eqref{eq:Atau}.
Then,
\begin{equation*}
\EE\left[z_{t,k} f\left(\xi^{\tau}_{t+\tau,j}\right)\right] = \EE\left[z_{t,k}\right]\EE\left[ f\left(\xi^{\tau}_{t+\tau,j}\right)\right]=0
\end{equation*}
Note that as $\beta^{f,j,\tau}_k = 0$, $\boldsymbol{\zeta}_k = \boldsymbol{\beta}^{f,j,\tau}_{-k}$. As $k$ is not a $\tau$-lagged ancestor, its $0$-lagged children and descendants cannot be either. Hence, their innovation terms cannot contribute to $\boldsymbol{\xi}_{t,-k}^{\top}\boldsymbol{\zeta}_k$ such that $z_{t,k} \perp \boldsymbol{\xi}_{t,-k}^{\top}\boldsymbol{\zeta}_k$ and $z_{t,k} \perp w_{t,k}$.

\subsection{Proof of Theorem \ref{SVAR:theo:norm}}
Throughout this proof, we apply the law of large numbers in various places. The justification is presented in Section \ref{app:NED}.

With the law of large numbers and the continuous mapping theorem, we get
\begin{align*}
\dfrac{1}{T}\mathbf{x}_{\_p+1}^\top \mathbf{x}_{\_p+1} \overset{\mathbb{P}}{\to} \boldsymbol{\Sigma^{\mathbf{x}}} & \implies \dfrac{1}{T}\mathbf{x}_{\_p+1,-k}^\top \mathbf{x}_{\_p+1, -k} \overset{\mathbb{P}}{\to} \boldsymbol{\Sigma^{\mathbf{x}}_{-k,-k}} \implies T \left(\mathbf{x}_{\_p+1, -k}^\top \mathbf{x}_{\_p+1, -k}\right)^{-1} \overset{\mathbb{P}}{\to} \left(\boldsymbol{\Sigma^{\mathbf{x}}_{-k,-k}}\right)^{-1} \\
& \implies T \left\Vert\left(\mathbf{x}_{\_p+1, -k}^\top \mathbf{x}_{\_p+1, -k}\right)^{-1}\right\Vert \overset{\mathbb{P}}{\to} \left\Vert\left(\boldsymbol{\Sigma^{\mathbf{x}}_{-k,-k}}\right)^{-1}\right\Vert = \mathcal{O}\left(1\right)
\end{align*}
For $\mathbf{x}_{\_p+1, -k}^\top \mathbf{z}_k/T$ we get a stronger result. Consider any entry
\begin{equation*}
\dfrac{1}{T} \sum_{t=1}^{T} x_{t\_p+1, l}z_{t,k},
\end{equation*}
where $l$ could also represent a time-lagged entry. Now for $t' > t$ consider the autocovariance\\
$\EE\left[x_{t\_p+1, l}z_{t,k} x_{t'\_p+1, l}z_{t',k}\right]$. As argued in \ref{app:proof:bols}, $z_{t',k}$ is a combination of innovations from time $t'$ independent of all previous times. Hence, a contribution to the autocovariance could only come from $\EE\left[x_{t\_p+1, l}z_{t,k} \right]\EE\left[x_{t'\_p+1, l}z_{t',k}\right]$. But this is $0$ by definition such that there is no time correlation. Combining this with the fourth moment assumption, we can apply the stronger result in Theorem 7.1.1 of \cite{brockwell2009time} leading to $\left\Vert \mathbf{x}_{\_p+1, -k}^\top \mathbf{z}_k\right\Vert =\mathcal{O}_p\left(\sqrt{T}\right)$. Analogously, as $\boldsymbol{\xi}^{\tau}_j$ has bounded time dependence $\left\Vert \mathbf{x}_{-\tau-1\_p}^\top \boldsymbol{\xi}^{\tau}_j\right\Vert =\mathcal{O}_p\left(\sqrt{T}\right)$. Let $\mathbf{a}_j = \left(\mathbf{A}_\tau\right)_j$ be the effect from $\mathbf{x}_{t-1\_p}$ on $x_{t+\tau,j}$ and $\hat{\mathbf{a}}_j$ its least squares estimate.
\begin{align*}
\left\Vert \mathbf{z}_k - \hat{\mathbf{z}}_k\right\Vert_2^2 &= \left\Vert P_{-k} \mathbf{z}_k \right\Vert_2^2 \leq \left\Vert \mathbf{z}_k^\top \mathbf{x}_{\_p+1, -k} \right\Vert_2 \left\Vert \left( \mathbf{x}_{\_p+1, -k}^\top \mathbf{x}_{\_p+1, -k}\right)^{-1} \right\Vert_2 \left\Vert \mathbf{x}_{\_p+1, -k}^\top \mathbf{z}_k \right\Vert_2 \\
&= \mathcal{O}_p\left(\sqrt{T}\right)\mathcal{O}_p\left(\dfrac{1}{T}\right)\mathcal{O}_p\left(\sqrt{T}\right) = \mathcal{O}_p\left(1\right)\\
\left\Vert \boldsymbol{\xi}^{\tau}_j - \hat{\boldsymbol{\xi}}^{\tau}_j\right\Vert_2^2 &= \left\Vert P_{\mathbf{x}_{-\tau}}\boldsymbol{\xi}^{\tau}_j\right\Vert_2^2 \leq \left\Vert \left(\boldsymbol{\xi}^{\tau}_j\right)^\top \mathbf{x}_{-\tau-1\_p} \right\Vert_2 \left\Vert\left(\mathbf{x}_{-\tau-1\_p} ^\top \mathbf{x}_{-\tau-1\_p} \right)^{-1}\right\Vert_2 \left\Vert \mathbf{x}_{-\tau-1\_p}^\top \boldsymbol{\xi}^{\tau}_j\right\Vert_2\\
&= \mathcal{O}_p\left(\sqrt{T}\right)\mathcal{O}_p\left(\dfrac{1}{T}\right)\mathcal{O}_p\left(\sqrt{T}\right) = \mathcal{O}_p\left(1\right)\\
\left\Vert \mathbf{a}_j - \hat{\mathbf{a}}_j\right\Vert_2 &= \left\Vert \left(\mathbf{x}_{-\tau-1\_p} ^\top \mathbf{x}_{-\tau-1\_p} \right)^{-1}\mathbf{x}_{-\tau-1\_p}^\top \boldsymbol{\xi}^{\tau}_j\right\Vert_2 \leq \left\Vert \left(\mathbf{x}_{-\tau-1\_p} ^\top \mathbf{x}_{-\tau-1\_p} \right)^{-1}\right\Vert_2 \left\Vert\mathbf{x}_{-\tau-1\_p}^\top \boldsymbol{\xi}^{\tau}_j\right\Vert_2\\
& = \mathcal{O}_p\left(\dfrac{1}{T}\right)\mathcal{O}_p\left(\sqrt{T}\right) = \mathcal{O}_p\left(\dfrac{1}{\sqrt{T}}\right)\\
\left\Vert \boldsymbol{\xi}^{\tau}_j - \hat{\boldsymbol{\xi}}^{\tau}_j\right\Vert_\infty & = \left\Vert \mathbf{x}_{-\tau-1\_p}\left(\mathbf{a}_j - \hat{\mathbf{a}}_j\right)\right\Vert_\infty \leq \left\Vert \mathbf{x}_{-\tau-1\_p}\right\Vert_\infty \left\Vert \mathbf{a}_j - \hat{\mathbf{a}}_j\right\Vert_1 \leq \left\Vert \mathbf{x}_{-\tau-1\_p}\right\Vert_\infty \sqrt{pd}\left\Vert \mathbf{a}_j - \hat{\mathbf{a}}_j\right\Vert_2 \\
& = {\scriptstyle \mathcal{O}}_p\left(1\right).
\end{align*}
We use $P_{-k}  = \mathbf{x}_{\_p+1, -k}\left( \mathbf{x}_{\_p+1, -k}^\top \mathbf{x}_{\_p+1, -k}\right)^{-1} \mathbf{x}_{\_p+1, -k}^\top$ in the first equality and the according decomposition for $P_{\mathbf{x}_{-\tau}}$ in the second equality. For matrices, $\Vert \cdot \Vert_2$ denotes the spectral norm. The last equality follows as with the fourth moment assumption the maximum grows no faster than $\mathcal{O}_p\left(T^{1/4}\right)$.

Assess the numerator of the least squares coefficient
\begin{align*}
\left\vert \mathbf{z}_k^\top f\left(\boldsymbol{\xi}_j^{\tau}\right)-\hat{\mathbf{z}}_k^\top f\left(\hat{\boldsymbol{\xi}}_j^{\tau}\right)\right\vert &\leq \left\vert \mathbf{z}_k^\top \left(f\left(\boldsymbol{\xi}_j^{\tau}\right)- f\left(\hat{\boldsymbol{\xi}}_j^{\tau}\right)\right)\right\vert + \left\vert \left(\mathbf{z}_k -\hat{\mathbf{z}}_k\right)^\top f\left(\boldsymbol{\xi}_j^{\tau}\right)\right\vert + \left\vert \left(\mathbf{z}_k -\hat{\mathbf{z}}_k\right)^\top \left(f\left(\boldsymbol{\xi}_j^{\tau}\right)-f\left(\hat{\boldsymbol{\xi}}_j^{\tau}\right)\right)\right\vert\\
&\leq \left\Vert \mathbf{z}_k \right\Vert_2 \left\Vert f\left(\boldsymbol{\xi}_j^{\tau}\right)-f\left(\hat{\boldsymbol{\xi}}_j^{\tau}\right)\right\Vert_2 + \left\Vert \mathbf{z}_k -\hat{\mathbf{z}}_k \right\Vert_2 \left\Vert f\left(\boldsymbol{\xi}_j^{\tau}\right)\right\Vert_2 + \left\Vert \mathbf{z}_k -\hat{\mathbf{z}}_k \right\Vert_2 \left\Vert f\left(\boldsymbol{\xi}_j^{\tau}\right)-f\left(\hat{\boldsymbol{\xi}}_j^{\tau}\right)\right\Vert_2
\end{align*}
By the moment assumption, $\left\Vert \mathbf{z}_k \right\Vert_2 = \mathcal{O}_p\left(\sqrt{T}\right)$ and $\left\Vert f\left(\boldsymbol{\xi}_j^{\tau}\right)\right\Vert_2= \mathcal{O}_p\left(\sqrt{T}\right)$. We consider the difference in the nonlinearities.
First, apply Taylor's theorem
\begin{equation*}
f\left(\hat{\boldsymbol{\xi}}_j^{\tau}\right)-f\left(\boldsymbol{\xi}_j^{\tau}\right)=\left(f'\left(\boldsymbol{\xi}_j^{\tau}\right)+h_1\left(\hat{\boldsymbol{\xi}}_j^{\tau}, \boldsymbol{\xi}_j^{\tau}\right)\right)\odot \left(\hat{\boldsymbol{\xi}}_j^{\tau} - \boldsymbol{\xi}_j^{\tau}\right),
\end{equation*}
where $h_1\left(\cdot\right)$ is the Peano form of the remainder. All functions are meant to be applied elementwise and $\odot$ denotes elementwise multiplication. For $\hat{\xi}_{t,j}^{\tau} \rightarrow \xi_{t,j}^{\tau}$, it holds $h_1\left(\hat{\xi}_{t,j}^{\tau}, \xi_{t,j}^{\tau}\right)\rightarrow 0$. Then,

\begin{align*}
\left\Vert f\left(\boldsymbol{\xi}_j^{\tau}\right)-f\left(\hat{\boldsymbol{\xi}}_j^{\tau}\right)\right\Vert_2^2 &= \sum_{t=1}^{T} \left(f\left(\xi_{t,j}^{\tau}\right)-f\left(\hat{\xi}_{t,j}^{\tau}\right)\right)^2 = \sum_{t=1}^{T} \left(f'\left(\xi_{t,j}^{\tau}\right) + h_1\left(\hat{\xi}_{t,j}^{\tau}, \xi_{t,j}^{\tau}\right)\right)^2\left(\xi_{t,j}^{\tau}-\hat{\xi}_{t,j}^{\tau}\right)^2\\
& \leq \left\Vert f'\left(\boldsymbol{\xi}_j^{\tau}\right)+h_1\left(\hat{\boldsymbol{\xi}}_j^{\tau}, \boldsymbol{\xi}_j^{\tau}\right)\right\Vert_\infty^2\sum_{t=1}^{T} \left(\xi_{t,j}^{\tau}-\hat{\xi}_{t,j}^{\tau}\right)^2\\
&=\left\Vert f'\left(\boldsymbol{\xi}_j^{\tau}\right)+h_1\left(\hat{\boldsymbol{\xi}}_j^{\tau}, \boldsymbol{\xi}_j^{\tau}\right)\right\Vert_\infty^2\left\Vert \boldsymbol{\xi}^{\tau}_j - \hat{\boldsymbol{\xi}}^{\tau}_j\right\Vert_2^2={\scriptstyle\mathcal{O}}_p\left(T\right).
\end{align*}
The maximum norm for $f'\left(\boldsymbol{\xi}_j^{\tau}\right)$ can be bounded at ${\scriptstyle\mathcal{O}}_p\left(\sqrt{T}\right)$ by the moment assumption, and that for $h_1\left(\hat{\boldsymbol{\xi}}_j^{\tau}, \boldsymbol{\xi}_j^{\tau}\right)$ is ${\scriptstyle\mathcal{O}}_p\left(1\right)$ by the properties of the remainder. In summary,
\begin{equation*}
\dfrac{1}{T}\hat{\mathbf{z}}_k^\top f\left(\hat{\boldsymbol{\xi}}_j^{\tau}\right) = \dfrac{1}{T}\mathbf{z}_k^\top f\left(\boldsymbol{\xi}_j^{\tau}\right) + {\scriptstyle \mathcal{O}}_p\left(1\right) = \EE\left[z_{t,k} f\left(\xi^{\tau}_{t+\tau,j}\right)\right]+ {\scriptstyle \mathcal{O}}_p\left(1\right).
\end{equation*}
Consider the denominator
\begin{align*}
\left\vert \mathbf{z}_k^\top \mathbf{z}_k - \hat{\mathbf{z}}_k^\top \hat{\mathbf{z}}_k \right\vert & = \left\vert \mathbf{z}_k^\top \mathbf{z}_k -  \mathbf{z}_k^\top P_{-k}^\perp \mathbf{z}_k\right\vert = \left\vert \mathbf{z}_k^\top \left(I - P_{-k}^\perp  \right) \mathbf{z}_k\right\vert = \left\Vert \mathbf{z}_k - \hat{\mathbf{z}}_k\right\Vert_2^2 = \mathcal{O}_p\left(1\right) \quad \text{such that} \\
\dfrac{1}{T}  \hat{\mathbf{z}}_k^\top \hat{\mathbf{z}}_k & = \dfrac{1}{T} \mathbf{z}_k^\top \mathbf{z}_k + {\scriptstyle \mathcal{O}}_p\left(1\right) =\EE\left[z_{t,k}^2\right] + {\scriptstyle \mathcal{O}}_p\left(1\right).
\end{align*}
Hence, $\hat{\beta}_k^{f,j,\tau}$ is indeed a consistent estimator.

For non-ancestors, we require a faster convergence for the numerator term. Due to orthogonality of the residuals and as $\beta_k^{f,j,\tau}=0$
\begin{equation*}
\hat{\mathbf{z}}_k^\top f\left(\hat{\boldsymbol{\xi}}_j^{\tau}\right) = \hat{\mathbf{z}}_k^\top \left( f\left(\hat{\boldsymbol{\xi}}_j^{\tau}\right) -  \hat{\boldsymbol{\xi}}^{0} \boldsymbol{\beta}^{f,j,\tau}\right).
\end{equation*}
Assess the approximation error of this term
\begin{align*}
&\left\vert \mathbf{z}_k^\top \left( f\left(\boldsymbol{\xi}_j^{\tau}\right) -  \boldsymbol{\xi}^{0} \boldsymbol{\beta}^{f,j,\tau}\right) - \hat{\mathbf{z}}_k^\top \left( f\left(\hat{\boldsymbol{\xi}}_j^{\tau}\right) -  \hat{\boldsymbol{\xi}}^{0} \boldsymbol{\beta}^{f,j,\tau}\right) \right\vert \leq \\
&\left\vert \mathbf{z}_k^\top \left(\left(f\left(\boldsymbol{\xi}_j^{\tau}\right) -  \boldsymbol{\xi}^{0} \boldsymbol{\beta}^{f,j,\tau}\right) - \left( f\left(\hat{\boldsymbol{\xi}}_j^{\tau}\right) -  \hat{\boldsymbol{\xi}}^{0} \boldsymbol{\beta}^{f,j,\tau}\right)\right) \right\vert +
\left\vert \left(\mathbf{z}_k -\hat{\mathbf{z}}_k\right)^\top \left( f\left(\boldsymbol{\xi}_j^{\tau}\right) -  \boldsymbol{\xi}^{0} \boldsymbol{\beta}^{f,j,\tau}\right)\right\vert + \\
&\left\vert \left(\mathbf{z}_k -\hat{\mathbf{z}}_k\right)^\top \left(\left(f\left(\boldsymbol{\xi}_j^{\tau}\right) -  \boldsymbol{\xi}^{0} \boldsymbol{\beta}^{f,j,\tau}\right) - \left( f\left(\hat{\boldsymbol{\xi}}_j^{\tau}\right) -  \hat{\boldsymbol{\xi}}^{0} \boldsymbol{\beta}^{f,j,\tau}\right)\right)\right\vert.
\end{align*}
The last term is controlled at ${\scriptstyle \mathcal{O}}_p\left(\sqrt{T}\right)$ with the Cauchy-Schwarz inequality using the rates from before. For the others, we make use of the structure of the process. Apply Taylor's theorem again
\begin{equation*}
f\left(\hat{\boldsymbol{\xi}}_j^{\tau}\right)-f\left(\boldsymbol{\xi}_j^{\tau}\right)=\left(f'\left(\boldsymbol{\xi}_j^{\tau}\right)+h_1\left(\hat{\boldsymbol{\xi}}_j^{\tau}, \boldsymbol{\xi}_j^{\tau}\right)\right)\odot \left(\hat{\boldsymbol{\xi}}_j^{\tau} - \boldsymbol{\xi}_j^{\tau}\right) = \left(f'\left(\boldsymbol{\xi}_j^{\tau}\right)+h_1\left(\hat{\boldsymbol{\xi}}_j^{\tau}, \boldsymbol{\xi}_j^{\tau}\right)\right)\odot  \left(\mathbf{x}_{-\tau-1\_p}\left(\mathbf{a}_j - \hat{\mathbf{a}}_j\right)\right),
\end{equation*}
Hence,
\begin{align*}
\left\vert \mathbf{z}_k^\top \left(f\left(\boldsymbol{\xi}_j^{\tau}\right) - f\left(\hat{\boldsymbol{\xi}}_j^{\tau}\right) \right) \right\vert &\leq \left\Vert \mathbf{z}_k^\top  \left(\left(f'\left(\boldsymbol{\xi}_j^{\tau}\right)+h_1\left(\hat{\boldsymbol{\xi}}_j^{\tau}, \boldsymbol{\xi}_j^{\tau}\right)\right)\odot  \mathbf{x}_{-\tau-1\_p}\right)\right\Vert_2 \left\Vert \mathbf{a}_j - \hat{\mathbf{a}}_j \right\Vert_2 \\
& \leq \left(\left\Vert \mathbf{z}_k^\top  \left(f'\left(\boldsymbol{\xi}_j^{\tau}\right)\odot  \mathbf{x}_{-\tau-1\_p}\right)\right\Vert_2 + \left\Vert \mathbf{z}_k^\top \left(h_1\left(\hat{\boldsymbol{\xi}}_j^{\tau}, \boldsymbol{\xi}_j^{\tau}\right)\odot  \mathbf{x}_{-\tau-1\_p}\right)\right\Vert_2 \right) \left\Vert \mathbf{a}_j - \hat{\mathbf{a}}_j \right\Vert_2.
\end{align*}
Note that these norms are over fixed dimensional vectors such that controlling one element of the vector is the same as controlling the norm. For the first summand, use $z_{t,k} \perp \xi_{t+\tau,j}^{\tau} , \mathbf{x}_{t-1\_p}$ such that the sum is over mean $0$ terms and hence ${\scriptstyle \mathcal{O}}_p\left(T\right)$. Consider any element in the second vector
\begin{equation*}
\left\vert\sum z_{t,k} h_1\left(\hat{\xi}_{t+\tau,j}^{\tau},\xi_{t+\tau,j}^{\tau}\right)x_{t-t',l}\right\vert \leq \left\Vert h_1\left(\hat{\boldsymbol{\xi}}_j^{\tau}, \boldsymbol{\xi}_j^{\tau}\right)\right\Vert_\infty \sum \left\vert z_{t,k} x_{t-t',l}\right\vert.
\end{equation*} 
As we control $\left\Vert \boldsymbol{\xi}^{\tau}_j - \hat{\boldsymbol{\xi}}^{\tau}_j\right\Vert_\infty$ the first factor is ${\scriptstyle \mathcal{O}}_p\left(1\right)$ while as the sum is $ \mathcal{O}_p\left(T\right)$ such that this term is controlled as ${\scriptstyle \mathcal{O}}_p\left(T\right)$ as well. The argument for 
\begin{equation*}
\left\vert \mathbf{z}_k^\top \left( \boldsymbol{\xi}^{0} \boldsymbol{\beta}^{f,j,\tau} -  \hat{\boldsymbol{\xi}}^{0} \boldsymbol{\beta}^{f,j,\tau}\right) \right\vert 
\end{equation*}
follows from the same principles without the remainder term in the Taylor expansion. As $\left\Vert \mathbf{a}_j - \hat{\mathbf{a}}_j \right\Vert_2 = \mathcal{O}_p\left(\dfrac{1}{\sqrt{T}}\right)$, these terms are ${\scriptstyle \mathcal{O}}_p\left(\sqrt{T}\right)$. It remains to look at the middle term.
\begin{align*}
&\left\vert \left(\mathbf{z}_k -\hat{\mathbf{z}}_k\right)^\top \left( f\left(\boldsymbol{\xi}_j^{\tau}\right) -  \boldsymbol{\xi}^{0} \boldsymbol{\beta}^{f,j,\tau}\right)\right\vert = \left\vert \mathbf{z}_k^\top P_{-k} \left( f\left(\boldsymbol{\xi}_j^{\tau}\right) -  \boldsymbol{\xi}^{0} \boldsymbol{\beta}^{f,j,\tau}\right)\right\vert \leq \\
&\left\Vert \mathbf{z}_k^\top \mathbf{x}_{\_p+1, -k} \right\Vert_2 \left\Vert \left( \mathbf{x}_{\_p+1, -k}^\top \mathbf{x}_{\_p+1, -k}\right)^{-1} \right\Vert_2 \left\Vert \mathbf{x}_{\_p+1, -k}^\top \left( f\left(\boldsymbol{\xi}_j^{\tau}\right) -  \boldsymbol{\xi}^{0} \boldsymbol{\beta}^{f,j,\tau}\right) \right\Vert_2.
\end{align*}
The first two factors are $\mathcal{O}_p\left(\sqrt{T}\right)\mathcal{O}_p\left(\dfrac{1}{T}\right)$ as before. In the last one, all sums are over mean $0$ terms. Columns of $\mathbf{x}_{\_p+1, -k}$ corresponding to time steps before $t$ are independent of $\boldsymbol{\xi}_j^{\tau}$ and $\boldsymbol{\xi}^{0}$. For the other columns, orthogonality is implied as $\boldsymbol{\beta}^{f,j,\tau}$ is the least squares coefficient. Thus, this factor is ${\scriptstyle \mathcal{O}}_p\left(T\right)$ and the term is ${\scriptstyle \mathcal{O}}_p\left(\sqrt{T}\right)$.
\begin{align*}
\dfrac{1}{\sqrt{T}}\hat{\mathbf{z}}_k^\top f\left(\hat{\boldsymbol{\xi}}_j^{\tau}\right) = \dfrac{1}{\sqrt{T}}\hat{\mathbf{z}}_k^\top \left( f\left(\hat{\boldsymbol{\xi}}_j^{\tau}\right) -  \hat{\boldsymbol{\xi}}^{0} \boldsymbol{\beta}^{f,j,\tau}\right) &= \dfrac{1}{\sqrt{T}}\mathbf{z}_k^\top \left( f\left(\boldsymbol{\xi}_j^{\tau}\right) -  \boldsymbol{\xi}^{0} \boldsymbol{\beta}^{f,j,\tau}\right)+{\scriptstyle \mathcal{O}}_p\left(1\right) = \dfrac{1}{\sqrt{T}}\mathbf{z}_k^\top \mathbf{w}_k+{\scriptstyle \mathcal{O}}_p\left(1\right)\\
&\overset{\mathbb{D}}{\to} \mathcal{N} \left\{0, \dfrac{1}{T}\EE\left[\left(\mathbf{z}_k^\top \mathbf{w}_k\right)^2\right]\right\}, 
\end{align*}
where we use the central limit theorem and Slutsky's theorem. By construction $\EE\left[z_{t,k}w_{t,k}\right]=0$. As $\mathbf{z}_k$, and $\mathbf{w}_k$ have time dependence only over a limited interval, the central limit theorem \citep[Theorem~6.4.2]{brockwell2009time} can be applied.
\begin{equation*}
\EE\left[\left(\mathbf{z}_k^\top \mathbf{w}_k\right)^2\right] = \EE\left[\left(\sum_t z_{t,k}w_{t,k}\right)^2\right]=\sum_t \sum_{t'} \EE\left[z_{t,k}w_{t,k} z_{t',k}w_{t',k}\right]
\end{equation*}
It holds $z_{t,k} \perp w_{t,k}$ and $z_{t,k}\perp z_{t',k}$ for $t \neq t'$ as there is no time-dependence. Also, $z_{t,k} \perp w_{t',k}$ for $t' > t$ as the innovation terms in $\xi_{t'+\tau,j}^{\tau}$ and $  \boldsymbol{\xi}_{t'}^{0}$ are from later time steps. Thus, for $t'>t$
\begin{equation*}
\EE\left[z_{t,k}w_{t,k} z_{t',k}w_{t',k}\right] = \EE\left[z_{t,k}\right]\EE\left[w_{t,k} z_{t',k}w_{t',k}\right]=0.
\end{equation*}
The argument for $t' < t$ is equivalent such that
\begin{equation*}
\EE\left[\left(\mathbf{z}_k^\top \mathbf{w}_k\right)^2\right] = \sum_t\EE\left[z^2_{t,k}w^2_{t,k} \right]=T \EE\left[z^2_{t,k}\right]\EE\left[w^2_{t,k} \right].
\end{equation*}
Plugging this in and using Slutsky's theorem again
\begin{equation}\label{SVAR:eq:bhat-conv}
\sqrt{T}\hat{\beta}_k^{f,j,\tau} \overset{\mathbb{D}}{\to} \mathcal{N} \left\{0, \dfrac{\EE\left[w_{t,k}^2\right]}{\EE\left[z_{t,k}^2 \right]}\right\}
\end{equation}
For the variance estimate, we use
\begin{equation*}
\left\Vert f\left(\hat{\boldsymbol{\xi}}_j^{\tau}\right) -  \hat{\boldsymbol{\xi}}^{0} \hat{\boldsymbol{\beta}}^{f,j,\tau}\right\Vert_2^2 = \left(f\left(\hat{\boldsymbol{\xi}}_j^{\tau}\right) -  \hat{\boldsymbol{\xi}}^{0} \hat{\boldsymbol{\beta}}^{f,j,\tau}\right)^\top \left(f\left(\hat{\boldsymbol{\xi}}_j^{\tau}\right) -  \hat{\boldsymbol{\xi}}^{0} \hat{\boldsymbol{\beta}}^{f,j,\tau}\right),
\end{equation*}
for which we have
\begin{align*}
& \left\vert \left(f\left(\boldsymbol{\xi}_j^{\tau}\right) -  \boldsymbol{\xi}^{0} \boldsymbol{\beta}^{f,j,\tau}\right)^\top \left(f\left(\boldsymbol{\xi}_j^{\tau}\right) -  \boldsymbol{\xi}^{0} \boldsymbol{\beta}^{f,j,\tau}\right) - \left(f\left(\hat{\boldsymbol{\xi}}_j^{\tau}\right) -  \hat{\boldsymbol{\xi}}^{0} \hat{\boldsymbol{\beta}}^{f,j,\tau}\right)^\top \left(f\left(\hat{\boldsymbol{\xi}}_j^{\tau}\right) -  \hat{\boldsymbol{\xi}}^{0} \hat{\boldsymbol{\beta}}^{f,j,\tau}\right)\right\vert \leq \\
& 2\left\Vert f\left(\boldsymbol{\xi}_j^{\tau}\right) -  \boldsymbol{\xi}^{0} \boldsymbol{\beta}^{f,j,\tau} \right\Vert_2 \left\Vert \left(f\left(\boldsymbol{\xi}_j^{\tau}\right) -  \boldsymbol{\xi}^{0} \boldsymbol{\beta}^{f,j,\tau}\right) - \left(f\left(\hat{\boldsymbol{\xi}}_j^{\tau}\right) -  \hat{\boldsymbol{\xi}}^{0} \hat{\boldsymbol{\beta}}^{f,j,\tau}\right)\right\Vert_2 + \\
&\left\Vert \left(f\left(\boldsymbol{\xi}_j^{\tau}\right) -  \boldsymbol{\xi}^{0} \boldsymbol{\beta}^{f,j,\tau}\right) - \left(f\left(\hat{\boldsymbol{\xi}}_j^{\tau}\right) -  \hat{\boldsymbol{\xi}}^{0} \hat{\boldsymbol{\beta}}^{f,j,\tau}\right)\right\Vert_2^2 ={\scriptstyle \mathcal{O}}_p\left(T\right) \quad \text{such that}\\
& \left\Vert f\left(\hat{\boldsymbol{\xi}}_j^{\tau}\right) -  \hat{\boldsymbol{\xi}}^{0} \hat{\boldsymbol{\beta}}^{f,j,\tau}\right\Vert_2^2 / T = \left\Vert f\left(\boldsymbol{\xi}_j^{\tau}\right) -  \boldsymbol{\xi}^{0} \boldsymbol{\beta}^{f,j,\tau}\right\Vert_2^2 / T + {\scriptstyle \mathcal{O}}_p\left(1\right)  = \EE\left[w_{t,k}^2\right] + {\scriptstyle \mathcal{O}}_p\left(1\right) = \mathcal{O}_p\left(1\right).
\end{align*}
The law of large numbers applies here as 
\begin{equation*}
\left\Vert f\left(\boldsymbol{\xi}_j^{\tau}\right) -  \boldsymbol{\xi}^{0} \boldsymbol{\beta}^{f,j,\tau}\right\Vert_2^2 / T
\end{equation*}
can be split into several converging sums of i.i.d.\ random variables.
Thus, we get
\begin{equation*}
\hat{\sigma}^2 = \mathcal{O}_p\left(1\right) \quad \text{and} \quad
\widehat{\text{var}}\left(\hat{\beta}^{f,j,\tau}_l\right)  =\mathcal{O}_p\left(\dfrac{1}{T}\right) \quad \forall l.
\end{equation*}
For non-ancestors $k$,
\begin{equation*}
\widehat{\text{var}}\left(\sqrt{T}\hat{\beta}^{f,j,\tau}_l\right) = T \widehat{\text{var}}\left(\hat{\beta}^{f,j,\tau}_l\right) \overset{\mathbb{P}}{\to} \dfrac{\EE\left[w_{t,k}^2\right]}{\EE\left[z_{t,k}^2 \right]}, 
\end{equation*}
i.e., the estimated variance approaches the asymptotic variance leading to the desired standard normal pivot.

\subsection{Near-epoch dependence}\label{app:NED}
We adapt the concept of near-epoch dependence, see, e.g., \cite{davidson1997strong, davidson2002establishing}. Define for $\tau \leq t$ $\mathfrak{F}_\tau^t=\sigma\left(\boldsymbol{\epsilon}_\tau, \ldots, \boldsymbol{\epsilon}_t\right)$ the $\sigma$-field generated by a subset of the innovation terms and $\EE_{t-m}^{t+m}\left[\cdot\right]$ the conditional expectation given $\mathfrak{F}_{t-m}^{t+m}$. Let $y_t$ be some random process.

\begin{definition}
$y_t$ is near-epoch dependent on $\left\{\boldsymbol{\epsilon}_t\right\}$ in $L_p$-norm, say $L_p$-NED, for $p>0$ if
\begin{equation*}
\EE\left[\left(y_t - \EE_{t-m}^{t+m}\left[y_t\right]\right)^p\right]^{1/p} \leq d_t \nu\left(m\right)
\end{equation*}
where $d_t$ is a sequence of positive constants, and $\nu\left(m\right) \overset{m \rightarrow \infty}{\rightarrow} 0$. It is said to be $L_p$-NED of size $-\mu$ if $\nu\left(m\right)= \mathcal{O}\left(m^{-\mu - \delta}\right)$ for some $\delta > 0$. It is said to be geometrically $L_p$-NED if $\nu\left(m\right)= \mathcal{O}\left(\exp\left(-\delta m\right)\right)$ for some $\delta > 0$.
\end{definition}
By establishing near-epoch dependence, we can apply the law of large numbers and the central limit theorem in appropriate places.

\begin{lemma}\label{lemm:xtj-NED}
Let $\mathbf{x}_t$ follow an SVAR \eqref{eq:svar} with finite second moments for which \ref{ass:iid} - \ref{ass:stable} holds. Then,
$x_{t,k}$ is geometrically $L_2$-NED on $\left\{\boldsymbol{\epsilon}_t\right\}$ for every $k \in \left\{1,\ldots\right\}$.
\end{lemma}
By the triangle inequality, the same holds then for every finite linear combination of $\mathbf{x}_t$.

\begin{lemma}\label{lemm:x2-NED}
Let $\mathbf{x}_t$ follow an SVAR \eqref{eq:svar} with finite second moments for which \ref{ass:iid} - \ref{ass:stable} holds. Then,
$x_{t,k}x_{t+\tau,l}$ is geometrically $L_1$-NED on $\left\{\boldsymbol{\epsilon}_t\right\}$ for every $l \in \left\{1,\ldots, d\right\}$ and $\tau < \infty$.
\end{lemma}

\begin{lemma}\label{lemm:xtxi-NED}
Let $\mathbf{x}_t$ follow an SVAR \eqref{eq:svar} with finite second moments for which \ref{ass:iid} - \ref{ass:stable} holds. Let $\xi_t$ be a quantity determined by $\mathfrak{F}_{t-\tau}^{t+\tau}$ for some $\tau < \infty$ such that $\EE\left[\xi_t\right]$ and $\EE\left[x_{t,k} \xi_t\right]$ are finite. Then,
$x_{t,k}\xi_t$ is geometrically $L_1$-NED on $\left\{\boldsymbol{\epsilon}_t\right\}$.
\end{lemma}

As the innovation sequence $\left\{\boldsymbol{\epsilon}_t\right\}$ is independent over time, it satisfies every mixing property. Thus, we can apply \cite{davidson1997strong}[Theorem 3.3] to obtain the LLN for $L_1$-NED quantities

\subsubsection{Proof of Lemma \ref{lemm:xtj-NED}}
Consider the form
\begin{equation*}
\mathbf{x}_{t\_p} = \tilde{\mathbf{B}} \mathbf{x}_{t-1\_p} + \boldsymbol{\xi}_{t\_p}=\sum_{\tau = 0}^{\infty} \tilde{\mathbf{B}}^\tau \boldsymbol{\xi}_{t-\tau\_p}
\end{equation*}
as in \cite{lutkepohl2005new} and \eqref{eq:comp}. Then
\begin{equation*}
\mathbf{x}_{t\_p} - \EE_{t-m}^{t+m}\left[\mathbf{x}_{t\_p}\right]=\mathbf{x}_{t\_p} - \EE_{t-m}^{t}\left[\mathbf{x}_{t\_p}\right]=\sum_{\tau = m+1}^{\infty} \tilde{\mathbf{B}}^\tau \boldsymbol{\xi}_{t-\tau\_p}.
\end{equation*}
Consider any $x_{t,k}=\mathbf{e}_k^\top \mathbf{x}_{t\_p}$, where $\mathbf{e}_k \in \mathbb{R}^{dp}$ is the according unit vector.
\begin{align*}
\EE\left[\left(x_{t,k} - \EE_{t-m}^{t}\left[x_{t,k}\right]\right)^2\right]&=\mathbf{e}_k^\top \EE\left[\sum_{\tau = m+1}^{\infty} \tilde{\mathbf{B}}^\tau \boldsymbol{\xi}_{t-\tau\_p} \left(\sum_{\tau' = m+1}^{\infty} \tilde{\mathbf{B}}^{\tau'} \boldsymbol{\xi}_{t-\tau'\_p}\right)^\top \right]\mathbf{e}_k \\
&=\mathbf{e}_k^\top \sum_{\tau = m+1}^{\infty} \tilde{\mathbf{B}}^\tau \EE\left[\boldsymbol{\xi}_{t\_p} \boldsymbol{\xi}_{t\_p}^\top\right] \left(\tilde{\mathbf{B}}^{\tau} \right)^\top \mathbf{e}_k\\
 & \leq  \sum_{\tau = m+1}^{\infty} \left\Vert \mathbf{e}_k^\top \tilde{\mathbf{B}}^\tau \right\Vert_2^2 \lambda_{\max} \left(\EE\left[\boldsymbol{\xi}_{t\_p} \boldsymbol{\xi}_{t\_p}^\top\right]\right) \leq C \sum_{\tau = m+1}^{\infty} \lambda_{\max}\left(\tilde{\mathbf{B}}\right)^{2 \tau},
\end{align*}
where $\lambda_{\max}\left(\cdot\right)$ denotes the largest absolute eigenvalue of a matrix, and $C$ is a accordingly chosen constant. Under the stability assumption \ref{ass:stable}, $\lambda_{\max}\left(\tilde{\mathbf{B}}\right) < 1$ such that the sum decreases as 
\begin{equation*}
\mathcal{O}\left(\lambda_{\max}\left(\tilde{\mathbf{B}}\right)\right)^{2m} \coloneqq\mathcal{O}\left( \exp\left(- 2\theta m\right)\right), \quad \text{where} \quad \theta = - \log\left(\lambda_{\max}\left(\tilde{\mathbf{B}}\right)\right) > 0.
\end{equation*}

\subsubsection{Proof of Lemma \ref{lemm:x2-NED}}
As $x_{t,k}$ is a linear combination of innovation terms, we can write
\begin{align*}
x_{t,k} &= \EE_{t-m}^{t+\tau}\left[x_{t,k}\right] + \EE_{-\infty}^{t-m-1}\left[x_{t,k}\right]\coloneqq \tilde{x}_{t,k} + \hat{x}_{t,k} \quad \text{and analogously}\\
x_{t+\tau,l} &= \EE_{t-m}^{t+\tau}\left[x_{t+\tau,l}\right] + \EE_{-\infty}^{t-m-1}\left[x_{t+\tau,l}\right]\coloneqq \tilde{x}_{t+\tau,l} + \hat{x}_{t+\tau,l}.
\end{align*}
Let $m\geq \tau$.
\begin{align*}
&\EE\left[\left\vert x_{t,k}x_{t+\tau,l} - \EE_{t-m}^{t+m}\left[x_{t,k}x_{t+\tau,l}\right]\right\vert\right] =\EE\left[\left\vert x_{t,k}x_{t+\tau,l} - \EE_{t-m}^{t+\tau}\left[x_{t,k}x_{t+\tau,l}\right]\right\vert\right]\\
=&\EE\left[\left\vert \left(\tilde{x}_{t,k} + \hat{x}_{t,k}\right)\left(\tilde{x}_{t+\tau,l} + \hat{x}_{t+\tau,l}\right)- \EE_{t-m}^{t+\tau}\left[\left(\tilde{x}_{t,k} + \hat{x}_{t,k}\right)\left(\tilde{x}_{t+\tau,l} + \hat{x}_{t+\tau,l}\right)\right]\right\vert\right]\\
= &\EE\left[\left\vert \tilde{x}_{t,k} \left(\hat{x}_{t+\tau,l} - \EE\left[\hat{x}_{t+\tau,l}\right]\right)+ \tilde{x}_{t+\tau,l} \left(\hat{x}_{t,k} - \EE\left[\hat{x}_{t,k}\right]\right)+\hat{x}_{t,k} \hat{x}_{t+\tau,l} - \EE\left[\hat{x}_{t,k} \hat{x}_{t+\tau,l}\right] \right\vert\right] \\
\leq & 2\left(\EE\left[\left\vert \tilde{x}_{t,k}\right\vert\right]\EE\left[\left\vert\hat{x}_{t+\tau,l}\right\vert\right]+\EE\left[\left\vert \tilde{x}_{t+\tau,l}\right\vert\right]\EE\left[\left\vert\hat{x}_{t,k}\right\vert\right] + \EE\left[\left\vert\hat{x}_{t,k} \hat{x}_{t+\tau,l}\right\vert\right]\right).
\end{align*}
As argued before $\hat{x}_{t,k}, \hat{x}_{t+\tau,l}$ have exponentially decreasing moments while as the moments of $\tilde{x}_{t,k}, \tilde{x}_{t+\tau,l}$ are bounded by the assumptions of the process. So, overall this sum is exponentially decreasing which establishes the Lemma.

\subsubsection{Proof of Lemma \ref{lemm:xtxi-NED}}
Decompose
\begin{equation*}
x_{t,k} = \tilde{x}_{t,k} + \hat{x}_{t,k}
\end{equation*}
as before. Let $m\geq \tau$.
\begin{align*}
&\EE\left[\left\vert x_{t,k}\xi_t - \EE_{t-m}^{t+m}\left[x_{t,k}\xi_t\right]\right\vert\right] = \EE\left[\left\vert \left(\tilde{x}_{t,k} + \hat{x}_{t,k}\right)\xi_t - \EE_{t-m}^{t+m}\left[\left(\tilde{x}_{t,k} + \hat{x}_{t,k}\right)\xi_t\right]\right\vert\right]\\
=&\EE\left[\left\vert \xi_t\left(\hat{x}_{t,k} - \EE_{t-m}^{t+m}\left[\hat{x}_{t,k}\right]\right) \right\vert\right]\leq 2 \EE\left[\left\vert\xi_t \right\vert\right]\EE\left[\left\vert\hat{x}_{t,k} \right\vert\right],
\end{align*}
which attains the exponential rate as argued before.

\subsection{Proof of Theorem \ref{SVAR:theo:adv}}
For simplicity, assume $\text{MA}^{\tau \rightarrow j}\left(k\right) = \text{MA}\left(k\right)$ such that
\begin{equation*}
z_{t,k} = x_{t,k} - \mathbf{x}_{t,\text{MA}^{\tau \rightarrow j}\left(k\right)}^\top \boldsymbol{\gamma}^{\tau \rightarrow j,k}.
\end{equation*}
We have 
\begin{equation*}
\beta_{k}^{f,j,\tau} = 0 \ \forall f\left(\cdot\right) \iff \EE\left[z_{t,k} f\left(\xi^{\tau}_{t+\tau,j}\right)\right] = \EE\left[\EE\left[z_{t,k} \mid \xi^{\tau}_{t+\tau,j}\right] f\left(\xi^{\tau}_{t+\tau,j}\right)\right] = 0 \ \forall f\left(\cdot\right) \iff \EE\left[z_{t,k} \mid \xi^{\tau}_{t+\tau,j}\right]=0.
\end{equation*}
Using the equality above, the last condition is equivalent to the one in the theorem. By construction 
\begin{equation*}
x_{t+\tau,j} =  x_{t,k} \alpha + \mathbf{x}_{t,\text{MA}^{\tau \rightarrow j}\left(k\right)}^\top \boldsymbol{\beta} + \tilde{\epsilon} \quad \text{for some } \alpha, \boldsymbol{\beta},
\end{equation*}
where $\tilde{\epsilon}$ is a linear combination of $x_{t',l}$ whose paths to $x_{t,k}$ are blocked by $\mathbf{x}_{t,\text{MA}^{\tau \rightarrow j}\left(k\right)}$ and noise terms $\epsilon_{t',l}$ with $t'>t$  such that
$x_{t,k} \perp \tilde{\epsilon} \mid \mathbf{x}_{t,\text{MA}^{\tau \rightarrow j}\left(k\right)}$. Hence, 
\begin{equation*}
x_{t,k} \not \perp x_{t+\tau,j} \mid \mathbf{x}_{t,\text{MA}^{\tau \rightarrow j}\left(k\right)} \implies \alpha \neq 0.
\end{equation*}
Then,
\begin{equation*}
\EE\left[z_{t,k} \xi^{\tau}_{t+\tau,j}\right] =\EE\left[z_{t,k} x_{t+\tau,j}\right] = \alpha \EE\left[z_{t,k}x_{t,k}\right] + \EE\left[z_{t,k} \mathbf{x}_{t,\text{MA}^{\tau \rightarrow j}\left(k\right)}^\top\right]\boldsymbol{\beta} + \EE\left[z_{t,k} \tilde{\epsilon}\right] = \alpha \EE\left[z_{t,k}^2\right] \neq 0,
\end{equation*}
i.e., the identity function leads to a non-zero regression coefficient if the conditional independence does not hold. The first equality holds as $z_{t,k}$ is independent from $\mathbf{x}_{t'}$ for $t'<t$. In the next expression, the middle summand vanishes by the construction of the regression residual $z_{t,k}$. For the last summand, note that all contributions from $\epsilon_{t',l}$ with $t'>t$ are independent of $z_{t,k}$ trivially, while as contributions of other $x_{t',l}$ must be uncorrelated from $z_{t,k}$ as it is a least squares residual.

For the last part, assume
$x_{t,k} \perp x_{t + \tau, j} \mid \mathbf{x}_{t,\text{MA}^{\tau \rightarrow j}\left(k\right)}$. This is equivalent to\\
$x_{t,k} \perp \xi^{\tau}_{t + \tau, j} \mid \mathbf{x}_{t,\text{MA}^{\tau \rightarrow j}\left(k\right)}$ as $\xi^{\tau}_{t + \tau, j} - x_{t + \tau, j}$ only depends on time before $t$ such that $x_{t,k}$ cannot depend on it given $\mathbf{x}_{t,\text{MA}^{\tau \rightarrow j}\left(k\right)}$. Then, it follows
\begin{align*}
\EE\left[z_{t,k} \mid \xi^{\tau}_{t + \tau, j}\right] &= \EE\left[x_{t,k} - \mathbf{x}_{t,\text{MA}^{\tau \rightarrow j}\left(k\right)}^\top \boldsymbol{\gamma}^{\tau \rightarrow j,k} \mid \xi^{\tau}_{t + \tau, j}\right]\\
&=\EE\left[\EE\left[x_{t,k} - \mathbf{x}_{t,\text{MA}^{\tau \rightarrow j}\left(k\right)}^\top \boldsymbol{\gamma}^{\tau \rightarrow j,k} \mid \mathbf{x}_{t,\text{MA}^{\tau \rightarrow j}\left(k\right)}, \xi^{\tau}_{t + \tau, j}\right]\mid \xi^{\tau}_{t + \tau, j}\right]\\
&=\EE\left[\EE\left[x_{t,k} \mid \mathbf{x}_{t,\text{MA}^{\tau \rightarrow j}\left(k\right)}\right]- \mathbf{x}_{t,\text{MA}^{\tau \rightarrow j}\left(k\right)}^\top \boldsymbol{\gamma}^{\tau \rightarrow j,k} \mid \xi^{\tau}_{t + \tau, j}\right]=0,
\end{align*}
where the second to last equality uses conditional independence and the last follows trivially from the assumption.

Finally, we can consider the general case where $\text{MA}^{\tau \rightarrow j}\left(k\right) \subseteq \text{MA}\left(k\right)$. Let $l \in \text{CH}^{0}\left(k\right) \setminus \text{AN}^\tau\left(j\right)$. We know that $\xi_{t,l}$ has a regression coefficient of $0$ by Theorem \ref{theo:bols}. Hence, removing it from the model cannot change the remaining least squares parameters. Now, after removing $x_{t,l}$, its parents do not contribute to $z_{t,k}$ unless they are in $\text{MA}^{\tau \rightarrow j}\left(k\right)$ due to one of the other conditions. Thus, removing these additionally cannot change $z_{t,k}$ and hence $\beta_k^{f,j,\tau}$. Therefore, it suffices to analyze with $\text{MA}^{\tau \rightarrow j}\left(k\right)$ only.

\subsection{Combined p-values}\label{app:comb}
The arguments presented here follow mainly on \cite{meinshausen2009p}. But, by noting that one should focus on the order statistics and not continuous quantiles, we slightly improve the penalty term for the combined p-value. Also, we omit here the possibility of ignoring the lowest p-values as there might be cases where only one should be non-uniform.

Let $x_{t,j}$ and $x_{t,k}$ be two of the observed time series and $p_k^{j,0}, \ldots p_k^{j,p}$ all p-values for potential effects from $k$ to $j$. Sort these p-values from lowest to largest, say $p_{k,\left(1\right)}^j, \ldots p_{k,\left(r\right)}^j$ where $r=p+1$. We get our combined p-value as 
\begin{equation*}
p_k^j = \underset{i \in \left\{1,\ldots,r\right\}}{\min} \dfrac{r}{i} p_{k,\left(i\right)}^j \sum_{i' = 1}^r \dfrac{1}{i' }.
\end{equation*}

\begin{proposition}
If $p_k^{j,0}, \ldots p_k^{j,p}$ are all $\text{Uniform}\left(0,1\right)$, $p_k^j$ is a valid p-value, i.e.,
\begin{equation*}
\PP\left(p_k^j \leq \alpha\right) \leq \alpha \ \forall \alpha \in \left(0,1\right).
\end{equation*}
\end{proposition}

\subsubsection*{Proof}
Define 
\begin{equation*}
\pi_k^j(u) = \dfrac{1}{r} \sum_{i=1}^r 1 \left\{p_{k,\left(i\right)}^j \leq u\right\}.
\end{equation*}
We have
\begin{equation*}
\pi_k^j\left(\alpha \dfrac{i}{r}\right) \geq \dfrac{i}{r} \iff p_{k,\left(i\right)}^j \dfrac{r}{i} \leq \alpha 
\end{equation*}

Let $U$ be a random variable in $\left[0,1\right]$ and consider
\begin{equation*}
\underset{i \in \left\{1,\ldots,r\right\}}{\max} \dfrac{1\left\{U \leq \alpha i/r\right\}}{i/r} = 
\begin{cases}
0 \qquad \qquad \qquad U > \alpha\\
\dfrac{r}{\lceil Ur / \alpha \rceil} \qquad \text{otherwise}.
\end{cases}
\end{equation*}
If $U$ is uniformly distributed we get the expectation 
\begin{equation*}
\EE\left[\underset{i \in \left\{1,\ldots,r\right\}}{\max} \dfrac{1\left\{U \leq \alpha i/r\right\}}{i/r}\right] = \alpha \sum_{i=1}^r \dfrac{1}{i}
\end{equation*}
as for each possible $i$, there is a segment of length $\alpha / r$ where $\lceil Ur / \alpha \rceil = i$.

Thus, if the individual p-values are uniform
\begin{equation*}
\EE\left[\underset{i \in \left\{1,\ldots,r\right\}}{\max} \dfrac{1}{r}\sum_{\tau=0}^p\dfrac{1\left\{p_k^{j,\tau} \leq \alpha i/r\right\}}{i/r}\right] \leq \dfrac{1}{r}\sum_{\tau=0}^p\EE\left[\underset{i \in \left\{1,\ldots,r\right\}}{\max} \dfrac{1\left\{p_k^{j,\tau} \leq \alpha i/r\right\}}{i/r}\right]= \alpha \sum_{i=1}^r \dfrac{1}{i}.
\end{equation*}
Apply the definition of $\pi_k^j\left(\cdot\right)$ and the Markov inequality.
\begin{align*}
\alpha \sum_{i=1}^r \dfrac{1}{i} &\geq \EE\left[\underset{i \in \left\{1,\ldots,r\right\}}{\max} \dfrac{1}{r}\sum_{\tau=0}^p\dfrac{1\left\{p_k^{j,\tau} \leq \alpha i/r\right\}}{i/r}\right] = \EE\left[\underset{i \in \left\{1,\ldots,r\right\}}{\max} \dfrac{\pi_k^{j} \left(\alpha i/r\right)}{i/r}\right] \geq \PP\left(\underset{i \in \left\{1,\ldots,r\right\}}{\max} \dfrac{\pi_k^{j} \left(\alpha i/r\right)}{i/r} \geq 1\right)\\
&=\PP\left(\underset{i \in \left\{1,\ldots,r\right\}}{\max} 1\left\{\pi_k^{j} \left(\alpha i/r\right) \geq i/r \right\} \geq 1\right) = \PP\left(\exists i \in \left\{1,\ldots,r\right\}: \quad \pi_k^{j} \left(\alpha i/r\right) \geq i/r \right)\\
& = \PP\left(\exists i \in \left\{1,\ldots,r\right\}: p_{k,\left(i\right)}^j \dfrac{r}{i} \leq \alpha \right) = \PP\left(\underset{i \in \left\{1,\ldots,r\right\}}{\min} p_{k,\left(i\right)}^j \dfrac{r}{i} \leq \alpha \right) = \PP\left(p_k^j \leq \alpha \sum_{i=1}^r \dfrac{1}{i} \right).
\end{align*}
As $\alpha$ is arbitrary in this argument, this establishes the super-uniformity of the p-value. The only place where uniformity of the individual $p_k^{j,\tau}$ is invoked is when calculating the expectation of a bounded function. Hence, if the individual p-values are asymptotically uniform, we obtain an asymptotic result for $p_k^j$.

\section{Simulation results}
\subsection{Details on the simulation setup}\label{app:sim}
We use the following distributions for the $\epsilon_{t,j}$: two $t_7$ distributions, two centered uniform distributions, a centered Laplace distribution with scale $1$, and a standard normal distribution. All distributions are normalized to have unit variance. For each simulation run with $6$ variables in Sections \ref{sim-x4} and \ref{network-sim}, we randomly permute the distributions to assign them to $\epsilon_{t,1}$ to $\epsilon_{t,6}$. For the simulation runs with more variables in Appendix \ref{app:high}, we randomly draw distributions with replacement from those $6$.

The edges $x_{t,k} \rightarrow x_{t,j}$ with $k<j$ are present, i.e., the entry $\left(\mathbf{B}_0\right)_{jk}$ is non-zero, with probability $0.2$ each such that an average of $3$ parental connections exists.

We assign preliminary edge weights uniformly in $\left[0.5, 1\right]$. These are further scaled such that for every $x_{t,j}$ which has instantaneous ancestors, the standard deviation of
\begin{equation*}
     \xi_{t,k} - \epsilon_{t,k}
\end{equation*}
is uniformly chosen from $\left[\surd 0.5, \surd 2\right]$ to control the signal-to-noise ratio.

To initialize the graph and the weights, we use the function \texttt{randomDAG} from the \texttt{R}-package \texttt{pcalg} \citep{kalisch2012causal} before applying our changes to the weights to enforce the constraints.

The entries in $\mathbf{B}_1$ are non-zero with probability $0.1$. If so, they are sampled uniformly with absolute value in $\left[0.2,0.8\right]$ and assigned a random sign with equal probabilities. If the maximum absolute eigenvalue of $\tilde{\mathbf{B}}$ would be larger than $0.95$, $\mathbf{B}_1$ is shrunken such that this absolute eigenvalue is $0.95$ to ensure stability.

We initiate the time series randomly and discard the first $10^4$ observations to ensure strict stationarity (approximately).

For the simulations with a hidden variable, we randomly sample a variable and exclude its time series from the data to estimate the models. 
This hidden variable is not counted to calculate the power. However, to obtain the true ancestors, it is considered, i.e., the ancestors of the unobserved variable are true ancestors of the unobserved variable's descendants.

\subsection{Simulations with more time series}\label{app:high}
We repeat the simulations in Section \ref{network-sim} with more time series, namely $d=10$ and $d=50$.

\begin{figure}[t!]
     \centering
     \begin{subfigure}[b]{0.8\textwidth}
         \centering
         \includegraphics[width=\textwidth]{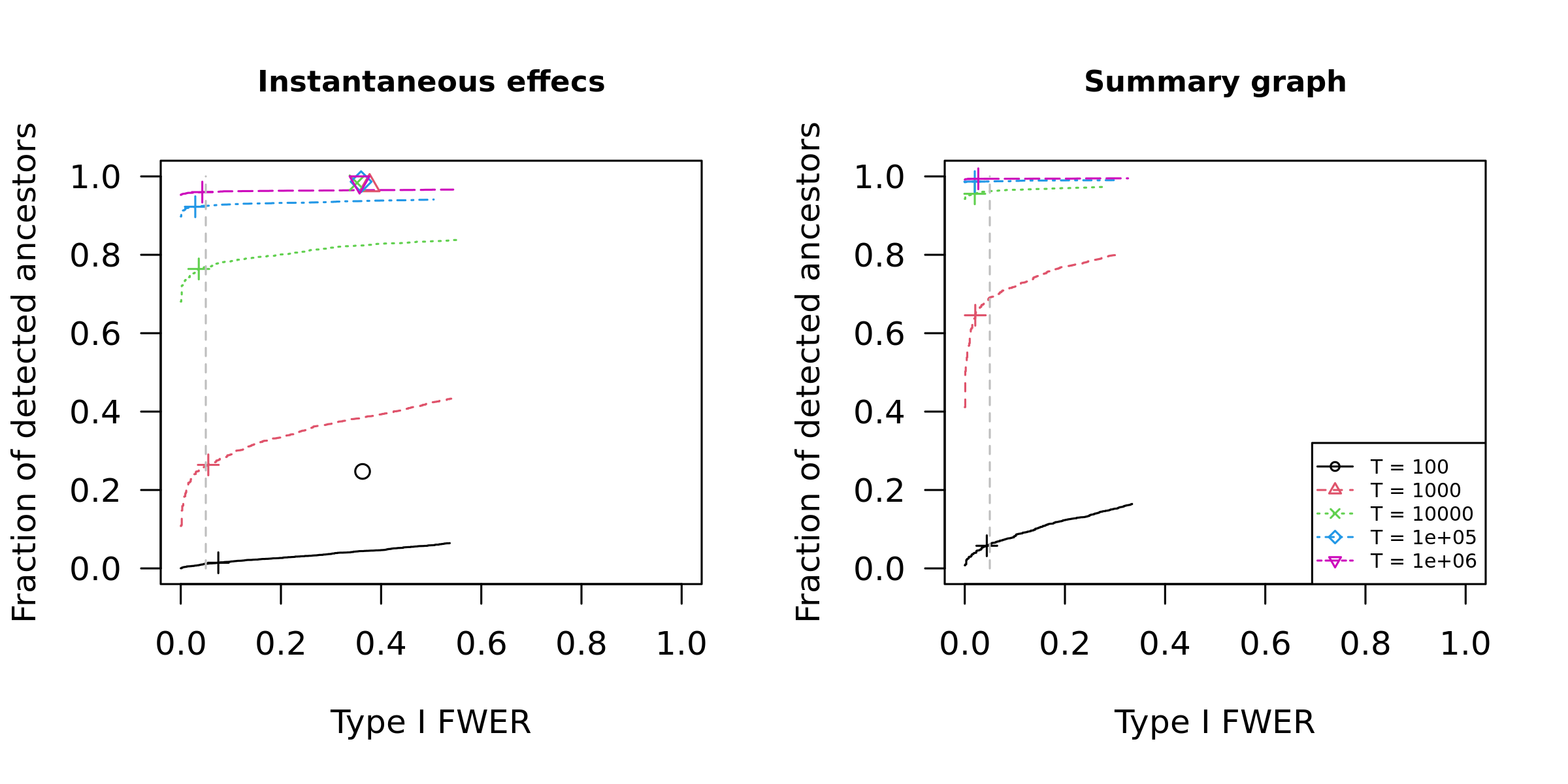}
         \caption{$d=10$}
     \end{subfigure}
     \hfill
     \begin{subfigure}[b]{0.8\textwidth}
         \centering
         \includegraphics[width=\textwidth]{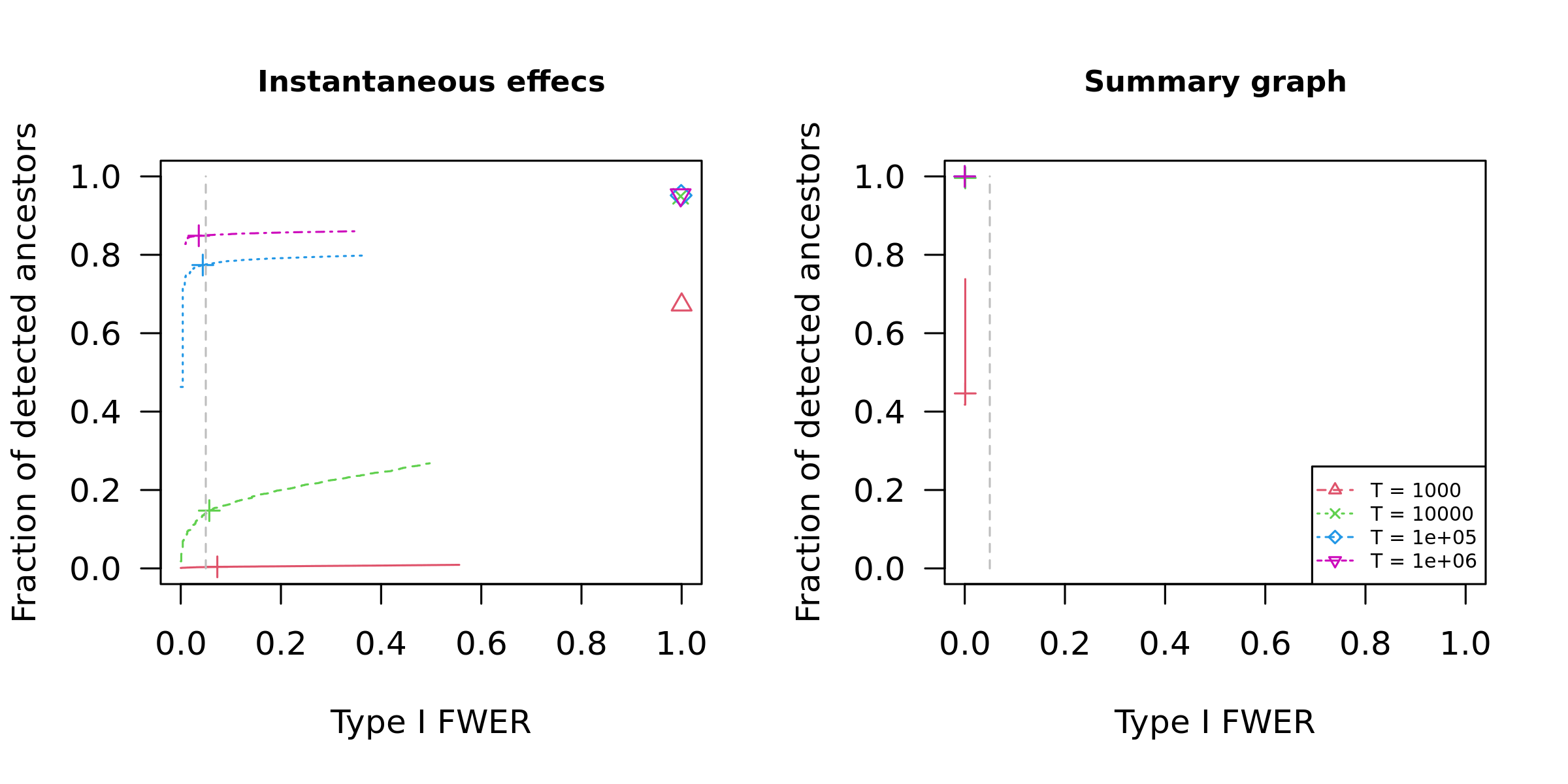}
         \caption{$d=50$}
     \end{subfigure}
     \caption{Nodewise ancestor detection in a structural vector autoregressive model of order $p=1$ with $d$ variables. See the caption of Figure \ref{SVAR:fig:graph-perf-rand} for explanations.}\label{fig:obs}
\end{figure}

In Figure \ref{fig:obs}, we show the performance when all time series are observed. For $d=50$, we cannot do estimation with only $100$ observations, so this is omitted in the simulation study. As we randomly sample from the distributions as outlined above, there is the possibility that several Gaussian distributions are assigned in a simulation run, and for $d=50$ this is very likely. Hence, LiNGAM has no consistency guarantee anymore, i.e., the error rate can remain high even for larger sample sizes, and ancestor regression may lose power to detect instantaneous effects. The Bonferroni-Holm correction together with the correction to obtain the summary p-values leads to a very high multiplier for $d=50$. Hence, the p-values for non-ancestors, which are approximately uniformly distributed, are mostly $1$ after correction, and the error rate is always low. Furthermore, there are not many true nulls for the summary effects which could lead to type I errors.

\begin{figure}[t!]
     \centering
     \begin{subfigure}[b]{0.8\textwidth}
         \centering
         \includegraphics[width=\textwidth]{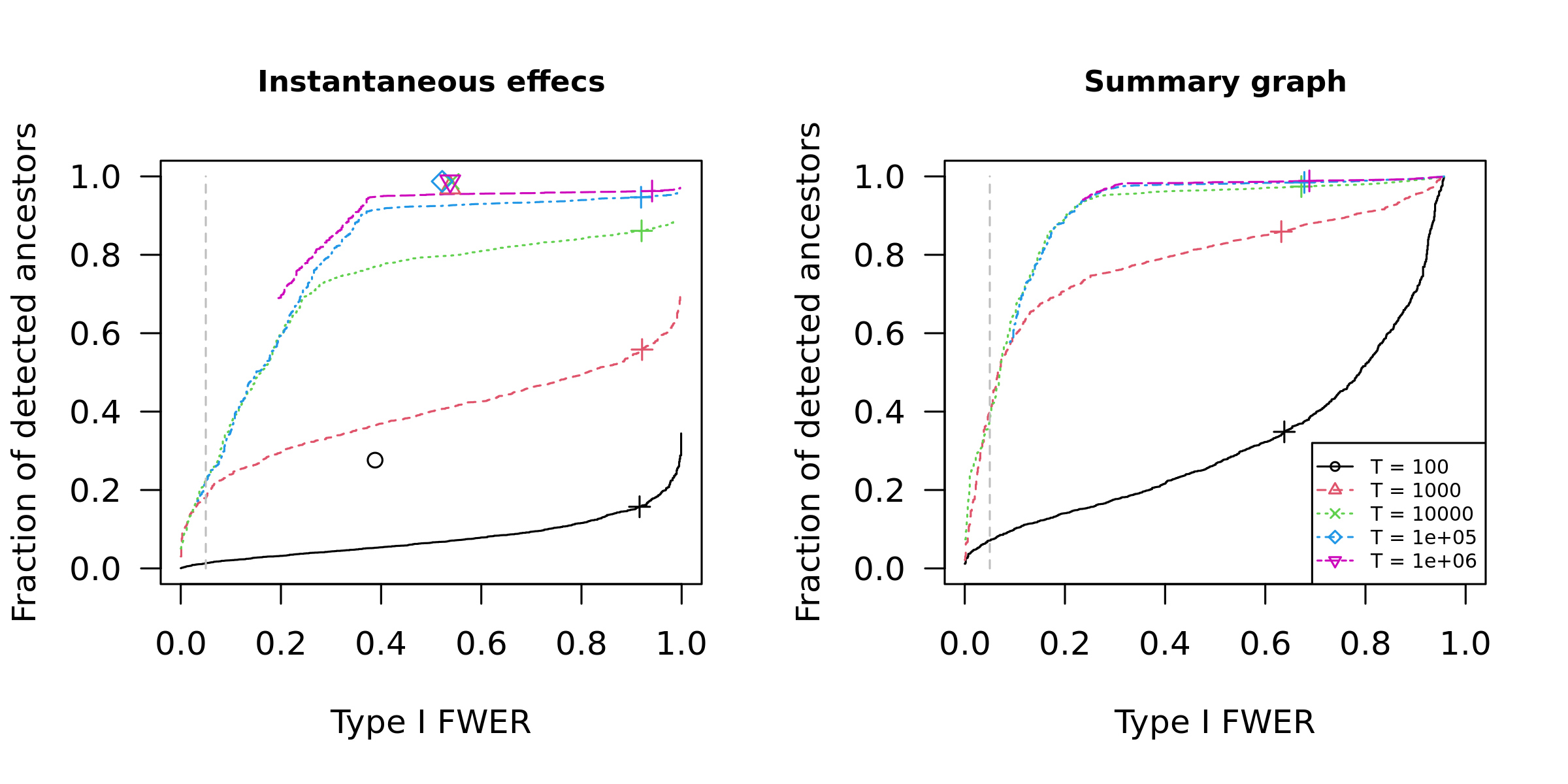}
         \caption{$d=10$}
     \end{subfigure}
     \hfill
     \begin{subfigure}[b]{0.8\textwidth}
         \centering
         \includegraphics[width=\textwidth]{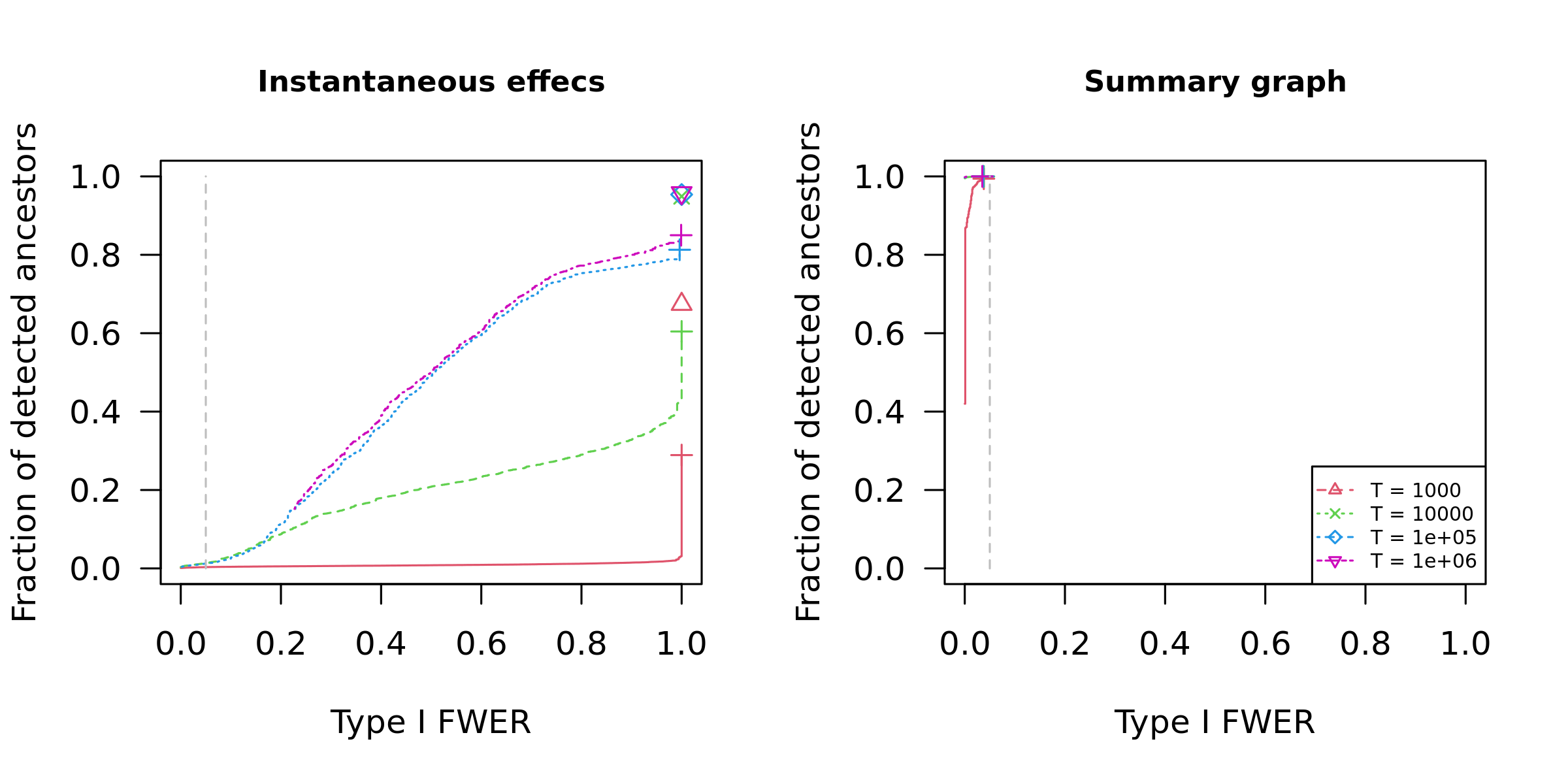}
         \caption{$d=50$}
     \end{subfigure}
     \caption{Nodewise ancestor detection in a structural vector autoregressive model of order $p=1$ with $d$ variables, whereof one is unobserved. See the caption of Figure \ref{SVAR:fig:graph-perf-rand} for explanations.}\label{fig:miss}
\end{figure}

The results when setting one time series at random as unobserved are shown in Figure \ref{fig:miss}. As for $d=6$, we lose error control when one time series is unobserved. There is still some separation between ancestors and non-ancestors, i.e., our test statistic provides an indication of what could be true ancestors. However, for $d=50$, it requires a very high sample size for reasonable performance. As outlined before, the potential for type I errors is lower for the summary effects. This applies even with an unobserved time series.
\end{document}